\documentclass[%
 reprint,
 amsmath,amssymb,
 aps,
]{revtex4-1}

\usepackage{graphicx}
\usepackage{dcolumn}
\usepackage{bm}

\usepackage{booktabs}
\usepackage{multirow}
\usepackage{threeparttable}
\usepackage{caption}

\begin{document}

\preprint{APS/123-QED}

\title{Evolution of amorphous carbon across densities: an inferential study}

\author{Bishal Bhattarai}
 \email{bb248213@ohio.edu}
\affiliation{Department of Physics and Astronomy, Condensed Matter and Surface Science Program (CMSS), Ohio University, Athens. Ohio 45701,USA }%

\author{Anup Pandey}
\email{ap439111@ohio.edu}
\affiliation{The Chemical and Engineering Materials Division (CEMD), Oak Ridge National Laboratory, Oak Ridge, TN, USA}

\author{D. A. Drabold}
\email{drabold@ohio.edu}
\affiliation{Department of Physics and Astronomy, Nanoscale and Quantum Phenomena Institute (NQPI), Ohio University, Athens, Ohio 45701, USA}%

\date{\today}

\begin{abstract}
In this paper, we offer large and realistic models of amorphous carbon spanning densities from 0.95 $g/cm^3$ to 3.5 $g/cm^3$. The models are \textit{designed} 
to agree as closely as possible with experimental diffraction data while simultaneously attaining a local minimum of a density functional Hamiltonian. The structure 
varies dramatically from interconnected wrapped and defective $sp^2$ sheets at 0.95 $g/cm^3$ to a nearly perfect tetrahedral topology at 3.5 $g/cm^3$. Force Enhanced Atomic Refinement (FEAR) was used and is
shown here to be computationally superior and more experimentally realistic than conventional \textit{ab initio}
melt quench methods. We thoroughly characterize our models by computing structural, electronic and vibrational spectra. The vibrational density of states 
of the 0.95 $g/cm^3$ model is strikingly similar to monolayer amorphous graphene. Our $sp^2$/$sp^3$ ratios are close to experimental predictions where 
available, a consequence of compelling a satisfactory fit for pair correlation function.

\end{abstract}

\maketitle


\section{\label{sec:level1}Introduction}
Amorphous materials are exploited for myriad applications such as thin-film transistors, solar photovoltaics, coatings and artificial 
heart-valves.\cite{DRMcKenzie1,DraboldEuro1,Roberston1} However, a lack of long range order in amorphous solids impose a challenge for a
condensed matter theorists. A logical approach for determining structure is to use experiment to infer structure. This is accepted practice for crystals, even those with extremely
large unit cells. For amorphous materials, a unique inversion is impossible because of the smooth structure factors and pair-correlation functions. The 
key shortcoming of such an approach is the unbiased inclusion of chemical information. A long used alternative: the method of ``melt quenching'' is limited 
by fast quenching rates and ignores a priori experimental information in the process of model formation.\cite{DraboldEuro1} We bridge the divide between 
these approaches in this paper.

For proper context we note that inverse modeling is experimentally driven where Reverse Monte Carlo (RMC)\cite{MCGreevyPusztai1} is used for modeling of
different amorphous systems.\cite{Partha1,Malley1,Keen1,GerebenPusztai1} RMC approach to match experimental information seems
logical and gives us computation time benefit.
Often these models result in highly-constrained or under-constrained structures, which may turn out 
be inaccurate or totally unrealistic.\cite{Opletal2,Jain1} To resolve these in-adequacies different experimentally motivated constraints have been proposed
such as: multiple scattering data\cite{JKWalters1,Hosokawa1,Gurman1},
bond-angle constraints \cite{Tucker1}, coordination constraints\cite{Malley1} and so on, which can be quite effective.\cite{Opletal2} 
The real concern about constraints is that they introduce bias into the modeling scheme.
Alternatively, energy functional based constraints involve minimization of total energy and total forces.
Numerous approaches\cite{Partha2,Opletal5,Anup1,Anup2,Anup3} depending upon stage for implementing minimization been explored using 
empirical/DFT interactions along with several other methods. \cite{Kiran1,Cliffe1,Cliffe2,Tersoff1,NAMarks1,TBMD2,LLi1}  

We have implemented Force Enhanced Atomic Refinement (\textbf{FEAR})\cite{Anup1,Anup2,Anup3} method in amorphous carbon (a-Carbon). FEAR has advantages 
over other contemporary inversion methods. It's ability to predict accurate structure with correct chemical composition, starting from a \textit{random} structure
without any constraints has been a feature of this approach. We have used FEAR with state of the 
art \textit{ab initio} interactions for our calculations.
FEAR has been tested in several materials, it is a robust and efficient method to model different amorphous systems.\cite{Anup1,Anup2}

In this paper, we present a series of models of a-Carbon at various densities using the same approach for all. We systematically report the dependence of observables on the density.
The paper is organized as follows, In section 2 we discuss the computational methodology. In section 3, we report our models 
and the methods of preparation.  
Section 4, mainly focuses on the structural properties of the models and comparisons to experiments. Section 5 
is devoted to the electronic properties of the system. Section 6 we describe the vibrational properties of these 
carbons. In section 7, we 
summarize our findings and discuss the effectiveness of
our approach by comparing it to the other known results.  
  
\section{\label{sec:level2} Methodology}

We have prepared four models of a-Carbon with 648 atoms at densities (3.50 $ g/cm^3 $, 2.99 $ g/cm^3 $, 2.44 $ g/cm^3 $ and 0.951 $ g/cm^3 $) using FEAR. 
In FEAR, we begin with \textit{randomly} chosen coordinates which are subjected to partial structural
refinement with ``M'' accepted RMC steps and partial relaxations with conjugate gradient (CG) method for ``N'' relaxation steps. This cycle is repeated until 
the model is fully converged (fitting the data and a minimum of the DFT interactions).\cite{Anup1,Anup2} To our knowledge, these are the largest \textit{ab initio} models offered to date for a-Carbon.

The relaxation step was performed with single-$\zeta$ basis, 
periodic boundary conditions and Harris functional at constant volume using \textbf{SIESTA}\footnote{DFT code using LDA with Ceperley Alder exchange correlation\cite{siesta}}.  As an additional check of our models, we have relaxed the converged models using 
the \textit{ab initio} package \textbf{VASP}\cite{Kresse2} with plane wave basis\cite{Kresse2}, 
 $\Gamma ({\vec{k}}=0)$, plane-wave cutoff of 400 eV and an energy convergence tolerance of $10^{-4}$ eV. 
To compare and contrast, we have also prepared \textit{ab initio} based MQ models. We have prepared three models (160 atom) each
using SIESTA (LDA, Harris functional) and VASP (LDA, self-consistency).\cite{Kresse1,Bloch1}

\section{\label{sec:level3} Models}

\begin{figure*}[!htb]
\minipage{0.40\textwidth}
  \includegraphics[width=1.00\linewidth]{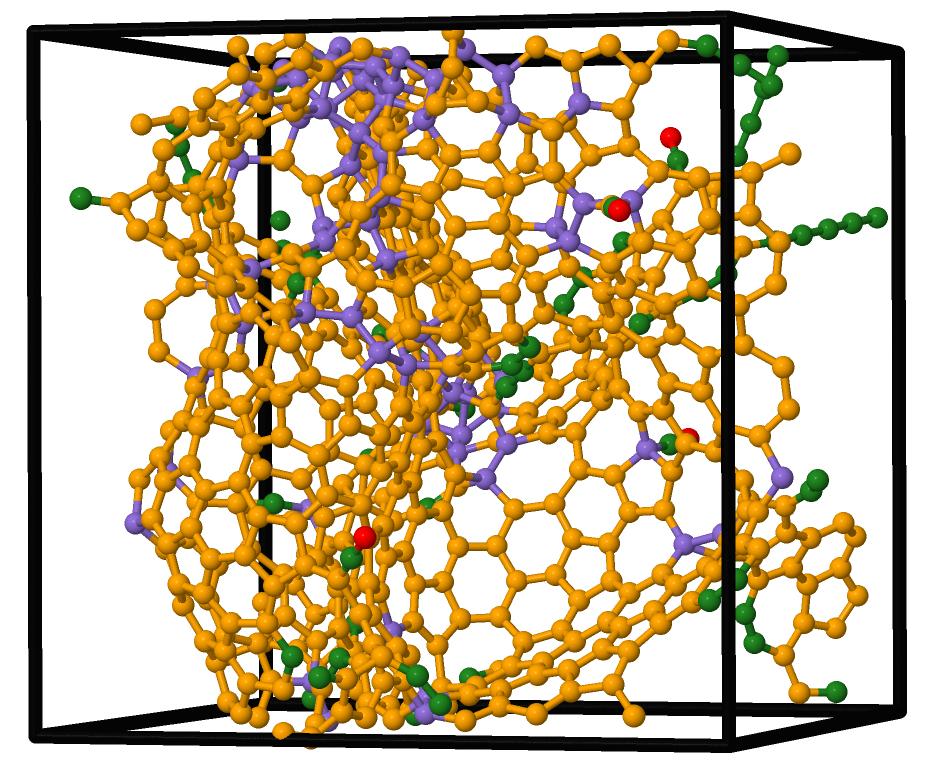}
  \caption*{ $\rho = 0.95 $  $g/cm^3$}
  \endminipage \hfill
\minipage{0.430\textwidth} 
  \includegraphics[width=1.00\linewidth]{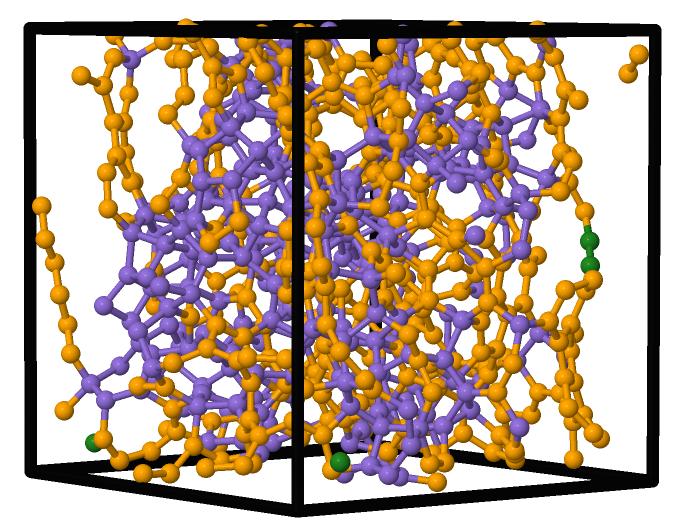}
  \caption*{ $\rho = 2.44 $  $g/cm^3$}
  \endminipage \hfill
\minipage{0.425\textwidth}%
  \includegraphics[width=1.00\linewidth]{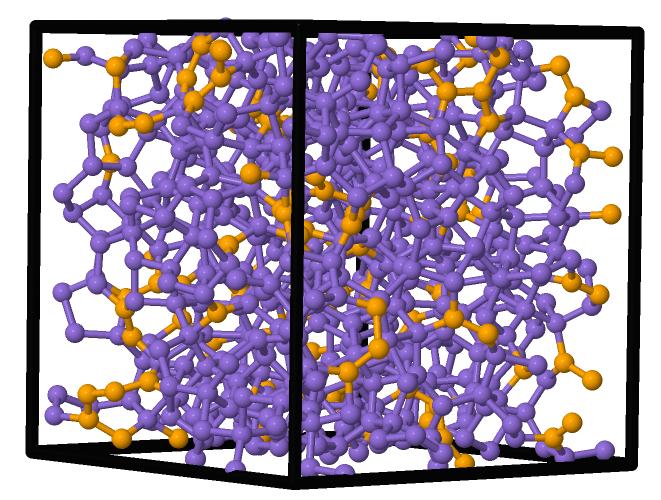}
  \caption*{ $\rho = 2.99 $  $g/cm^3$}
  \endminipage \hfill 
\minipage{0.40\textwidth}%
  \includegraphics[width=1.00\linewidth]{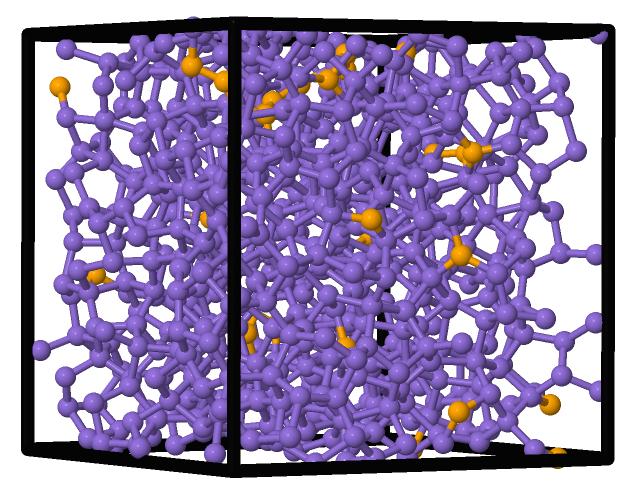}
  \caption*{ $\rho = 3.50 $  $g/cm^3$} 
  \endminipage 
\caption{(Color online) Visualization of the different bonding in amorphous carbon (F648): purple ($sp^3$), orange ($sp^2$), green ($sp$) and red (singly bonded).
\textit{Periodic boundary condition were used, only atoms in reference cell are shown.}
}
\end{figure*}

The starting random configuration is fitted to appropriate experimental data with \textbf{RMCProfile}\footnote{ RMC based applications for the structural refinement\cite{Tucker2}}. After every $\sim$100 accepted RMC moves\cite{Anup1,Anup2,Anup3}, the total energy and 
forces were evaluated (using a single force call) and the atoms were moved along the gradient to reduce the total energy.
We have chosen a maximum RMC step size of 0.25$\mathring{A}$-0.375$\mathring{A}$, a minimum approach of 1.05$\mathring{A}$-1.20 $\mathring{A}$,
with a fixed spacing of 0.02 $\mathring{A}$ and $0.04-0.085$ 
weight of the experimental data. Meanwhile, relaxation (CG) is carried out in SIESTA using a force tolerance of 
0.01 eV/$\mathring{A}$ and maximum CG displacement of 0.70 $\mathring{A}$.

In the meantime, we implemented MQ calculations with random 
coordinates, which  were  equilibrated at 7000 K, then cooled to 300 K, further equilibrated at 300 K and finally relaxed using CG method. 
This process employed a time step of 1.0 fs for a total time of 26 ps. We have also prepared a self-consistent MQ model using VASP. These 
models were started from random, then heated to 8000 K, equilibrated at 8000 K, cooled to 300 K and finally relaxed with CG method. A time step 
of 1.5 fs was used for total time of 24 ps.

These models will hereafter be identified as (F648, S160 and V160). The assigned nomenclature indicates: method of preparation (FEAR-SIESTA-VASP) and
number of atoms in the cell of each model. We have used our previous VASP prepared model (V72 at 0.92 $g/cm^3$)\cite{Bhattarai2} to 
compare the result of our lowest density model. Our models are summarized in Table. 1.

\begin{table*}[!ht]
\renewcommand*{\thefootnote}{\alph{footnote}}
\normalfont
\caption{\label{tab:table3} Nomenclature and details of our models: 
Position of first minimum of RDF ($r_{min}$), average co-ordination number (n), percentage of 
$sp^3$, $sp^2$ and $sp$ 
and total CPU time for the simulation ($T_0$).}\vspace{0.05cm}
\setlength{\arrayrulewidth}{0.45mm}
\begin{center}
\begin{tabular}{|c|c|c|c|c|c|c|c|c|c|c|c|}
\hline
\rule{0pt}{4mm}
 
\multirow{1}{12mm}{Models} & \multicolumn{3}{c|}{$\rho$ = 3.50 $(g/cm^3)$}& \multicolumn{3}{c|}{$\rho$ = 2.99 $(g/cm^3)$}& \multicolumn{3}{c|}{$\rho$ = 2.44 $(g/cm^3)$}& \multicolumn{2}{c|}{$\rho$ = 0.95 $(g/cm^3)$} \\ \cline{2-12}
\rule{0pt}{5mm}
\setcounter{footnote}{0}
&F648 & V160 & S160 & F648 & V160 & S160 & F648 & V160 & S160 & F648 & V72\footnotemark \\
\hline 

 
 \rule{0pt}{4 mm} 
 
n & 3.96 & 3.98 & 3.94  & {3.83} & {3.75}  & 3.85 & 3.41 & {3.26} &{3.58}&3.00& 2.67\\
\rule{0pt}{4 mm}

$\%$ of $sp^3$ & 96.00 & 97.50 & 93.75 & {82.70} & {75.00}  & 85.00 & 42.00 & {26.87} &{58.13}&10.80&---  \\
\rule{0pt}{4 mm}

$\%$ of $sp^2$ & 4.00 & 2.50 & 6.25  & {17.30} & {25.00}  & 15.00 & 57.40 & {72.50} &{41.25}&79.00&66.67  \\
\rule{0pt}{4 mm}

$\%$ of $sp$ & --- & --- & ---  & {---} & {---}  & --- & 0.60 & {0.63} &{0.62}&9.60&33.33  \\

\rule{0pt}{5 mm}


$T_0$\footnotemark & {28.12} & 100 & ---  & {30.73} & {100}  & --- & {23.58} & {100} &{---}&{---}& ---  \\

\hline 
\end{tabular}
\footnotesize 
\begin{tablenotes}
      \item[a]{\hspace{0.5cm} \textit{$^{a}$ at density $0.92$  $g/cm^3$.\cite{Bhattarai2}}}
      \item[b]{\hspace{0.5cm} \textit{$^{b}$  CPU time for fixed number of total cores.}}
    \end{tablenotes}    
\end{center}
 
\end{table*}

\setcounter{footnote}{2}

\begin{figure*}[!ht]
\begin{minipage}[b]{.475\textwidth}
  \centering
  \includegraphics[width=1.0\linewidth]{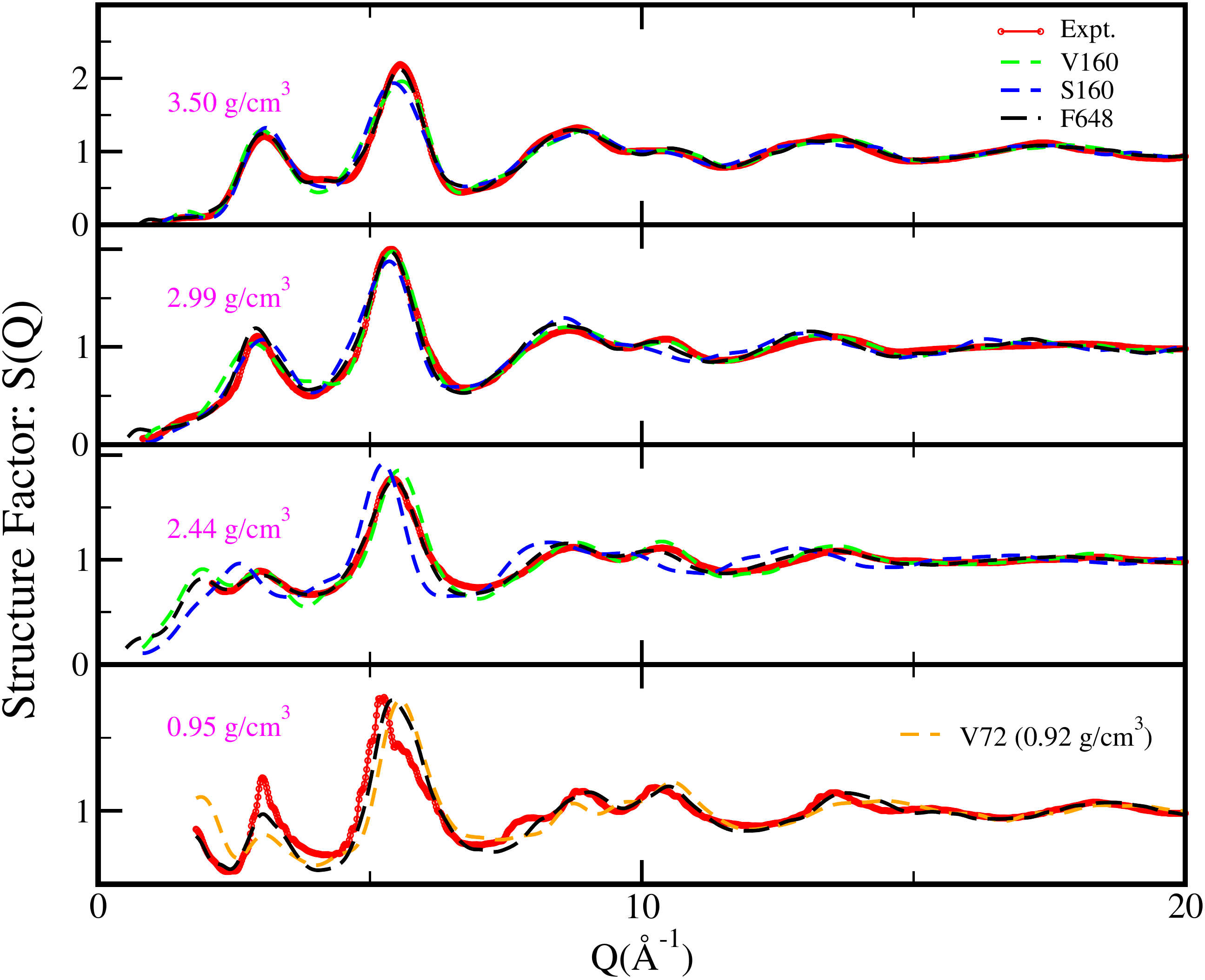}
\end{minipage}\hspace{0.2cm}
\begin{minipage}[b]{.475\textwidth}
  \centering
  \includegraphics[width=1.10\linewidth]{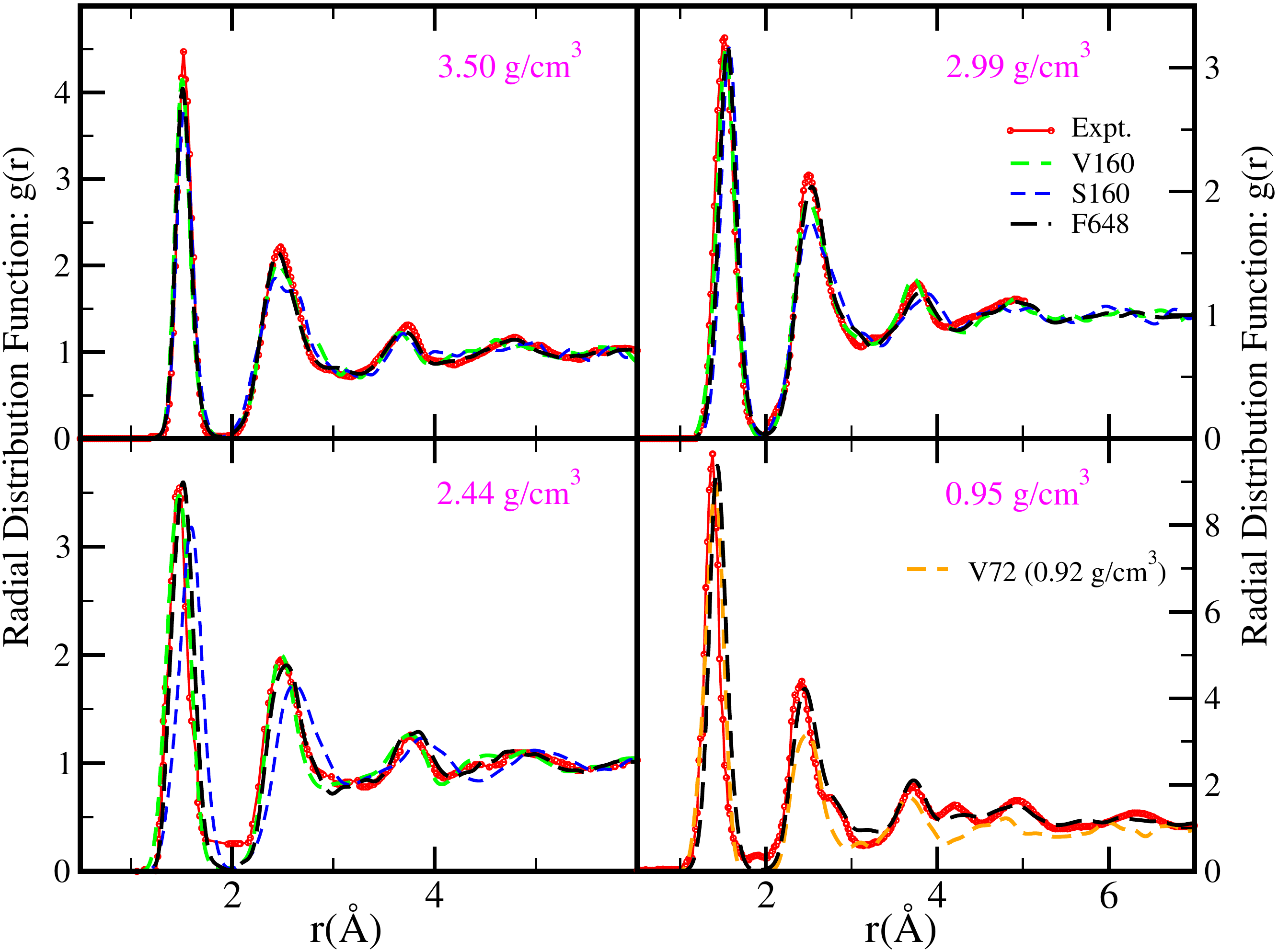}
\end{minipage}\vspace{0.1cm}

\caption{(Color online) \textbf{(Left panel)} Structure factor for different models and their comparison with experiments (or WWW model for  $3.50$ $g/cm^3$). \textbf{(Right panel)} 
  Radial distribution function of different models and their comparison with experiments. The experimental data are excerpted from previous literature.
  \cite{Opletal1,WWW1,Gilkes1,FangLi1,Opletal4}}
\label{fig:fig10231209894}
\end{figure*}

\section{\label{Structurefactorandpair}Structural Properties}

Structurally, amorphous carbon at density $3.50$ $g/cm^3$ is 
diamond-like ($sp^3$) bonded whereas near graphitic density $2.27$ $g/cm^3$ it is mostly $sp^2$ bonded and further at low densities
($<$ $2.0$ $g/cm^3$) we observe a few $sp$ conformation with mostly $sp^2$ bonded carbon structures.\cite{McCulloch1} This change in bonding preferences with density is 
shown in Fig. 1. We have assigned different color codes (via. Jmol \footnote{Jmol, an open-source Java viewer for chemical structures in 3D}) for varying bonding and it reveals at high densities $sp^2$ mainly inter-connects $sp^3$ 
networks while at low densities it is exactly vice-versa. Our $sp^2/sp^3$ ratios are close to experimental findings.\cite{ACFerrari1,PJFAllon1} In Fig. 2, we show a comparison of experimental static structure factor ($\textit{S(Q)}$) and radial distribution function (RDF, $\textit{g(r)}$) with our FEAR models.

At density $3.50$ $g/cm^3$, we have used a Wooten-Wearie-Winer (WWW)\cite{WWW1} model
as our input experimental diffraction data as no data is available for this density. The WWW model is obtained with bond-switching-algorithm\cite{WWW2} 
with perfect (100 $\%$ $sp^3$) bonding and has been an ideal model for tetrahedral amorphous systems\cite{TafenC1}.
We have close agreement for both $\textit{S(Q)}$ and $\textit{g(r)}$ with the WWW model, further we reproduce 96 $\%$ of the $sp^3$ content in our model (Table 1).
In contrast, earlier finding\cite{Wang2} report a lower concentration of $sp^3$ at this density.

We have used experimental diffraction data as our RMC input for next three calculations. At a density $2.99$ $g/cm^3$, we employed the neutron diffraction 
data of \textit{Gilkes et al.}\cite{Gilkes1} which is estimated to have 84$\%$ $sp^3$ bonding and a coordination-number (n) of 3.84. Our obtained $\textit{S(Q)}$ and $\textit{g(r)}$ are in an 
excellent agreement with experiment and we reproduce 82.70 $\%$ $sp^3$ bonding with a coordination(n) of 3.83. Similarly, at 
density $2.44$ $g/cm^3$, we have used experimental diffraction data of \textit{Li and Lannin}\cite{FangLi1} as our RMC input. We again obtain good 
agreement with the experimental diffraction data ($\textit{S(Q)}$ and $\textit{g(r)}$) while some deviations are seen in MQ models. 
These results for densities ($2.99$ $g/cm^3$ and $2.44$ $g/cm^3$) are in better agreement with the experiment compared to some earlier work.\cite{McCulloch1, Wang2,Opletal1}

Finally, at density $0.95$ $g/cm^3$, we employed neutron diffraction data obtained for silicon carbide-derived nanoporous carbon(SIC-CDC)\cite{Opletal4}
as FEAR input. Amorphous carbon at this density is also known as glassy carbon, a bit of a misnomer as the materials are not conventional glasses. The uncertainty of density, structure and significant H-content
makes it difficult to study glassy carbons.\cite{Malley1,Opletal3,Mildner1,Zetterstrom1,Jain1} 
Most calculations include strong assumptions, such as choosing a perfect graphitic or graphene sheets, co-ordination
restrictions, bond-angle restrictions and so on.\cite{Malley1,Opletal3,Pikunic1,Zetterstrom1} Some of these constrained models were found 
to be unstable and were subjected to change upon relaxation.\cite{Jain1} Additionally, accurate  \textit{``ab initio''}(complete basis DFT) based calculations of these glassy 
carbons have been limited to a few hundred atoms until this work. To check this significant inferences we repeated the calculation with a different random starting point and 
received a statistically equivalent model.

Our F648 model depicts the best extant picture of glassy carbon starting from random and without any bias.
We have a credible agreement with the experiment as
we have successfully reproduced major RDF peaks for glassy carbon  occurring at, 1.42 $\mathring{A}$, 2.46 $\mathring{A}$, 2.84 $\mathring{A}$ and so on.\cite{Parker1}
There is a slight deviations in the low Q range which was also observed in previous work\cite{Malley1,Opletal4}
done with ($>\textit{3200}$ atoms), concluding that it's not a finite size limitation.

\subsection{\label{AngleRing} Bond Angle Distribution and Ring statistics}

\begin{figure*}[!ht]
\begin{minipage}[b]{.475\textwidth}
  \centering
  \includegraphics[width=1.02\linewidth]{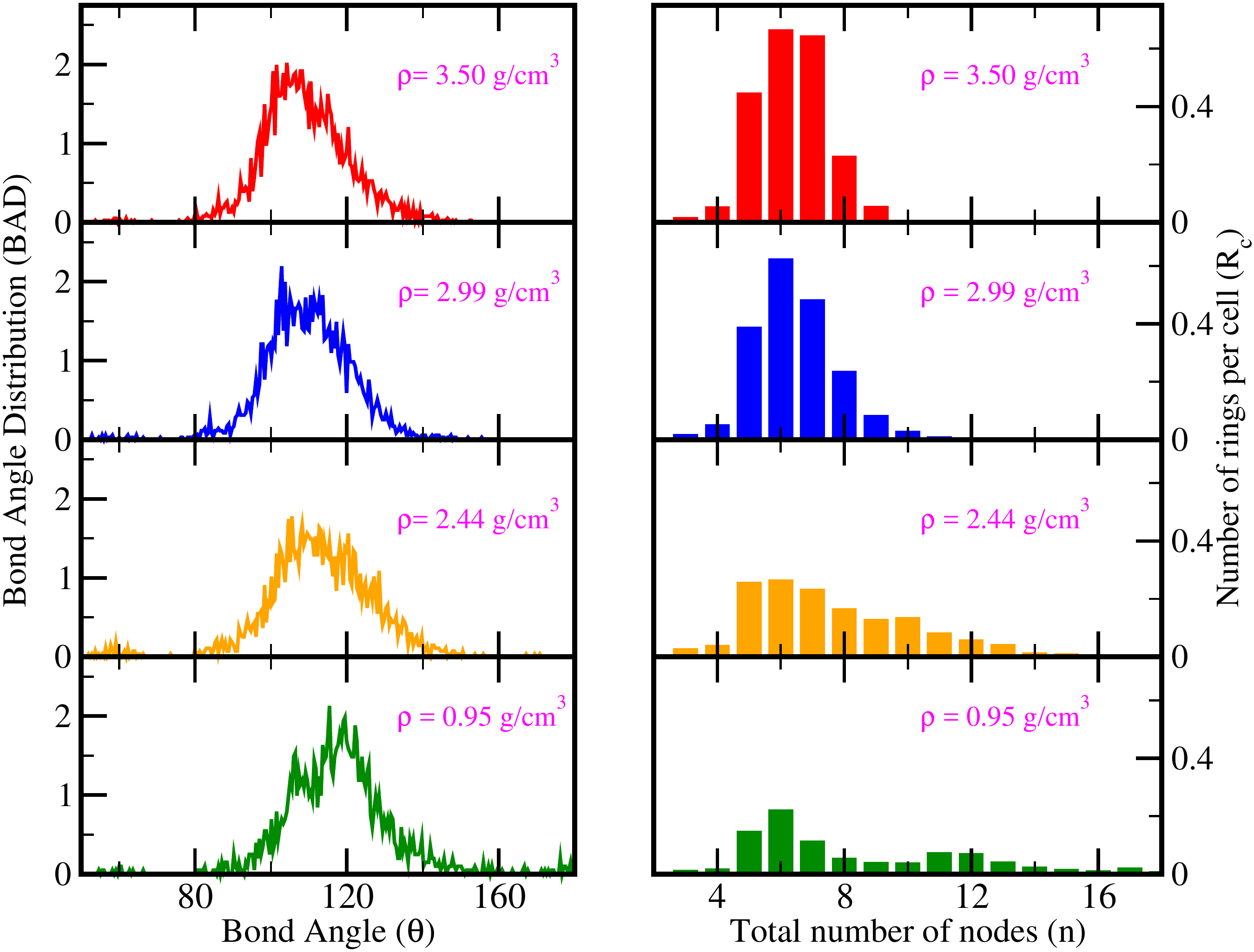}
\end{minipage}\hspace{0.2cm}
\begin{minipage}[b]{.475\textwidth}
 \centering
  \includegraphics[width=1.04\linewidth]{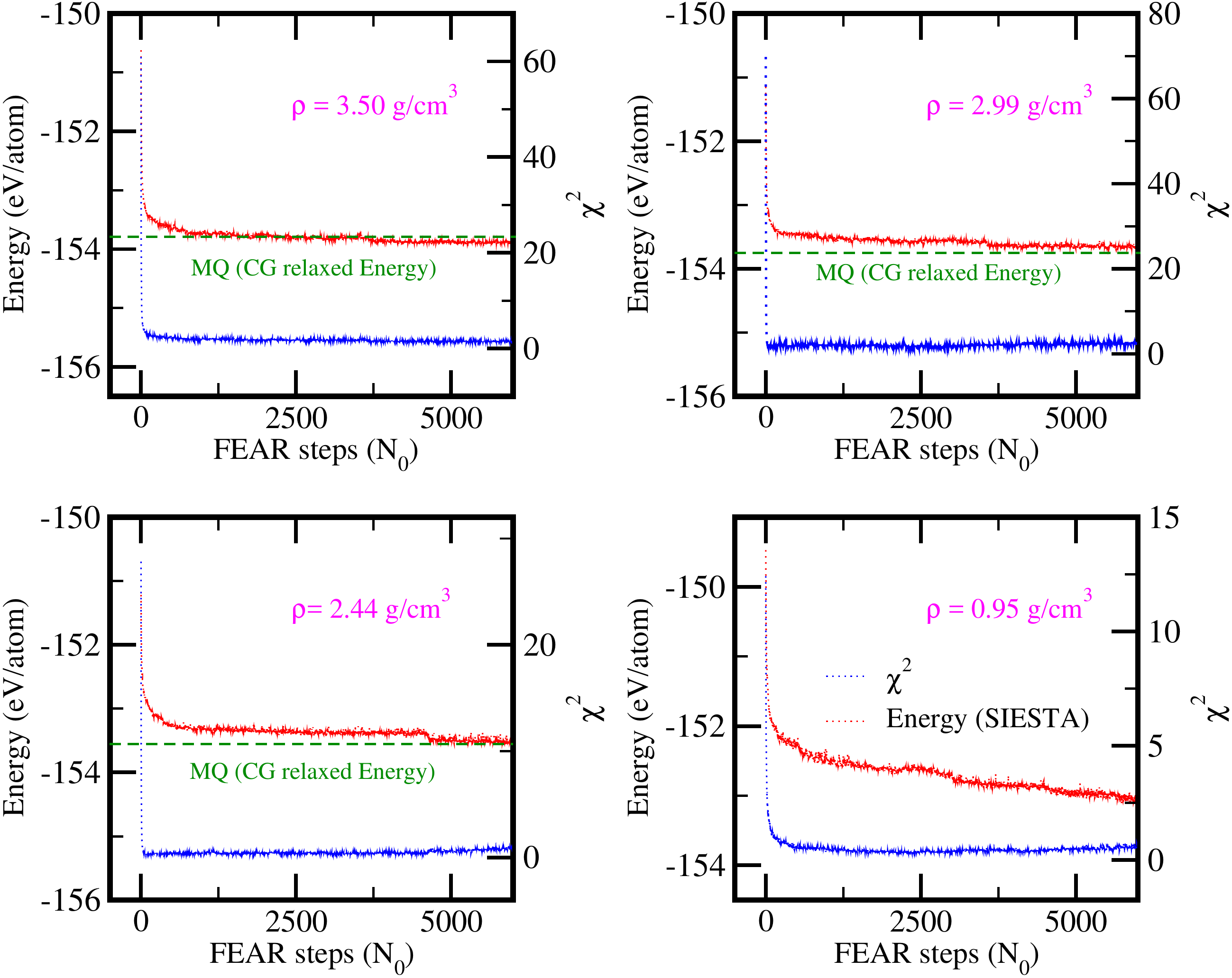}
\end{minipage}\vspace{0.1cm}

\caption{(Color online) \textbf{Left panel}) Bond angle distribution (BAD) and Ring statistics of F648 models. \textbf{(Right panel)} Plot of total energy (SIESTA) per atom (red line, F648) and cost function (blue line, F648) ($\chi^2$) versus number of FEAR steps ($N_0$).
 Final relaxed MQ energies are shown for comparison (green line, S160).}
\label{fig:fig1023124009}
\end{figure*}

Bond angle distribution (BAD) and ring statistics provide vital information about microstructure. In a typical RMC simulation with a perfect 
fit to experiment, a peak near $\sim 60.0^{o}$ is observed in BAD\cite{Anup1}. This is one of the major drawbacks of using RMC which FEAR
avoids. Although, constraints have been suggested\cite{Opletal1} to avoid these unrealistic cases, FEAR achieves it without external bias. 
We have reported our result for BAD and ring statistics in Fig. 3.

 At the high density the BAD peak is close to the tetrahedral angle of
$109.5^{o}$, with small deviation. At low densities the BAD peak is closer to $120.0^{o}$, indicating trigonal symmetry is dominant in these structures. It 
is reported that even with high $sp^2$ content BAD peak at low density is close to $117.0^{o}$.\cite{Beeman1} 

We have shown in Fig. 3 that amorphous carbons mostly prefer 5-7 membered ring structures. This is also true for the high $sp^2$ concentration 
structures which further clarifies that these a-C structures are different from graphite (only 6 membered rings). A negligible fraction of smaller 
ring structures were also observed but these are less than MD and other calculations.\cite{Opletal1} The ring statistics were evaluated with 
King's shortest path method\cite{King1} using ISAACS software.\cite{ISSACS1}

\subsection{ Convergence and stability of FEAR carbon}
In FEAR, we obtain low values of $\chi^2$ in conjunction with a local energy minimum\footnote{$\chi^2$ measures goodness of fit between experimental and FEAR model\cite{Anup1,Anup2,Anup3}}.
Our plot of variation of total energy (E) and $\chi^2$ is shown in Fig. 3. The results obtained shows that a initial structure in formed within few hundred 
FEAR steps where the system has attained the energy landscape for a-Carbon with some defects. These states have more or less the same average energy and 
as we move along with FEAR steps these defects are removed, thus leading to a chemically realistic structure.

\section{Electronic Density of States}

The concentration of $sp$, $sp^2$ and $sp^3$ states strongly influence the nature of the electronic density of states (EDOS).
As in the case of diamond, a-Carbon with high $sp^3$ is non-conducting.
We have presented plots of the EDOS of our F648 models in Fig. 4, where we have also decomposed the total EDOS by $sp^3$, $sp^2$ and $sp$
contributions. We can clearly see that that the $sp^2$ states for density $3.50$ $g/cm^3$ act as a defect and leads to formation of a pseudo-gap.\cite{DraboldC2} 
Subsequently, a-Carbon models at lower density are conducting as expected.\cite{Robertson1, McCulloch1}
In Fig. 4 (Right panel) we show the plot of Inverse Participation Ratio (IPR)\cite{Chen1}, IPR  gives information about the spatial localization of electronic states.
As seen in Fig. 4, the gap states for high density are highly localized while the lower two density have much more extended states.
This supports our observation that low density a-Carbon are conducting.

\begin{figure}[!ht]
  \centering
   \includegraphics[width=9.0 cm,]{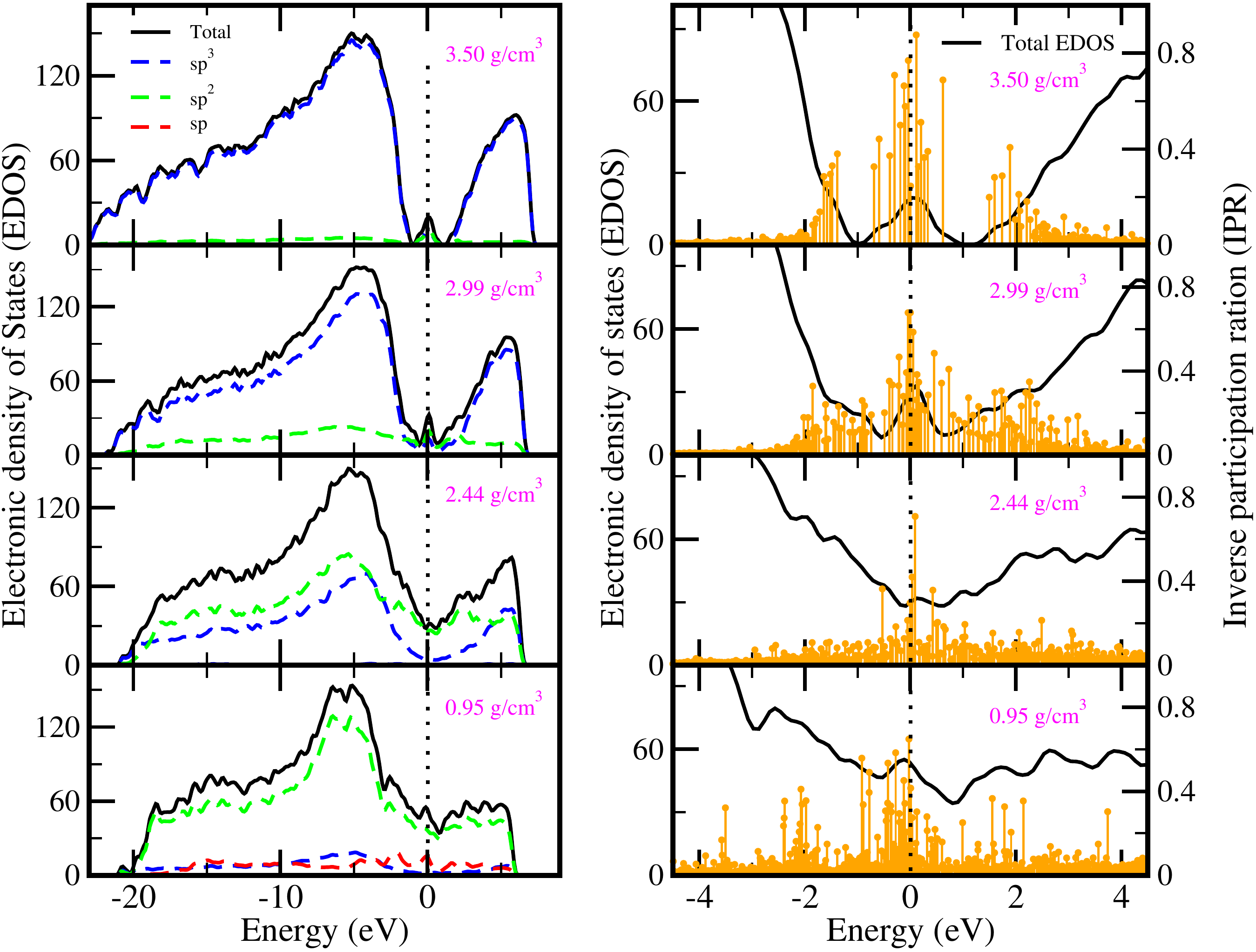}
  \caption{(Color online) Plot of EDOS (E$_{F}$=0 eV) for F648 models: (\textbf{Left panel}) black-solid (total EDOS), blue-dashed ($sp^3$ EDOS),
  green-dashed ($sp^2$ EDOS) and (green-dashed, $sp$ EDOS). (\textbf{Right panel})
  orange-drop lines(IPR) and black-solid (total EDOS).}\label{401}
\end{figure}

\section{\label{sec:level5} Vibrational Properties}

The vibrational density of states (VDOS) provides crucial information about changes in local bonding environment which is very effective
test for theoretical models\cite{Lopinski1} and offers a remarkably direct comparison between experiment and theory. It is well know that a-Carbon exhibits 
two major peaks in VDOS and Raman spectra show these occurring at: $\sim 1500$ $cm^{-1}$ and $\sim 800$ $cm^{-1}$.\cite{FangLi1}
In contrast, several theoretical models show a single broad peak occurring roughly at $\sim 1100$ $cm^{-1}$.\cite{DraboldC2,Wang3} 

We have calculated the vibrational density of states (VDOS) of our four F648 models. The dynamical matrix was obtained
by displacing each atom in 6-directions ($\pm x$,$\pm y$,$\pm z$) by a small displacement of 0.015 $\mathring{A}$ (see details\cite{Bhattarai1}). We have used Harris functional to 
our advantage for accelerating these computationally intensive calculations. Our VDOS plot for the four models are shown in Fig.5.

\begin{figure}[!ht]
  \centering
  \includegraphics[width=8.5 cm,]{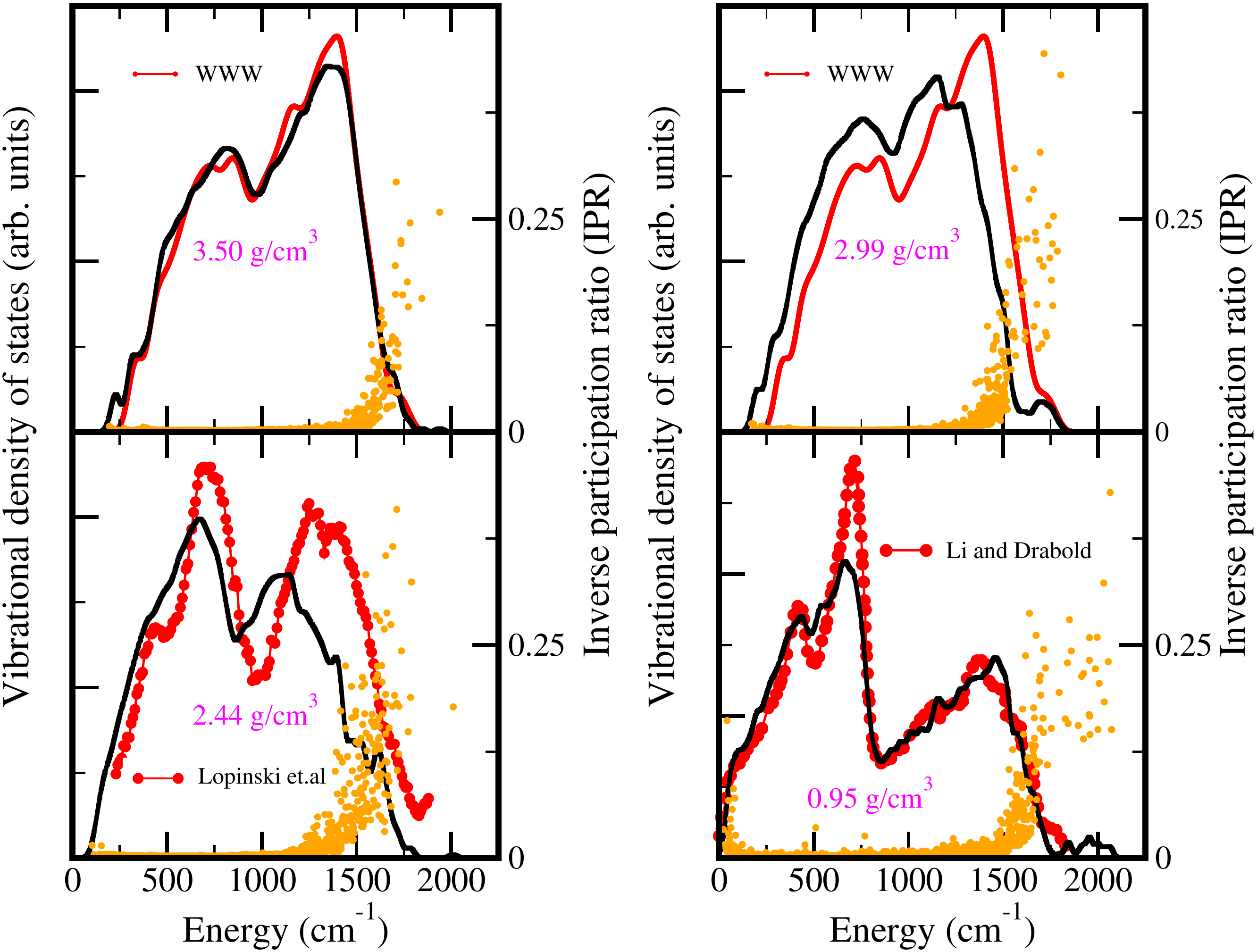}
  \caption{(Color online) Plot of Vibrational density of states (VDOS)(black line, F648), comparison with previous literatures(red dots and lines)\cite{WWW1,Lopinski1,Yuting1} and Inverse participation
  ratio (IPR) (orange dots) for F648 models.} 
\end{figure}

Our results show reasonable agreement with the literature. There is distinct bifurcation seen in our F648 models as seen in several experiments.
At $3.50$ $g/cm^3$, we compare our result with VDOS obtained for 216-WWW model.\cite{WWW1}
We observed a slight shift for $2.99$ $g/cm^3$ as compared to model at $3.50$ $g/cm^3$. At $2.44$ $g/cm^3$, we have compared our results ($sp^3$ fraction $42.0\%$) with 
experimental data\cite{Lopinski1} obtained for amorphous carbon containing $sp^3$ fraction at $60\%\pm 10\%$, we have a qualitative match with the 
experimental finding. The position of two peaks and their relative intensity is reported to slightly differ for different incident energies and $sp^3$ fraction.\cite{Lopinski1,Papanek1}

At low density $0.95$ $g/cm^3$, our model resembles distorted graphene structures (see Fig.1). We have 
compared our results with 2D a-graphene result of \textit{Li and Drabold}\cite{Yuting1}. The plots bear a remarkable similarity, most notably the peak 
occurring at $\sim 700 cm^{-1}$ and $\sim 1400 cm^{-1}$. This is surprising in view of the two models have different topology (one is 3-D, the other one 2-D). We have also computed the inverse participation ratio (IPR) for our F648 models. IPR, gives localization of these 
vibrational modes.\cite{Bhattarai1,Bhattarai2} Our obtained result for IPR shows that vibrational modes are extended at the low frequency regime and 
localized modes are only observed a higher frequency than $\sim 1500 cm^{-1}$, which are likely to be localized stretching modes.\cite{Wang3,Bhattarai2}

\section{\label{Comparison} Conclusions} 

We have used a \textit{uniform approach} to model a-Carbon using FEAR at various densities. We have used method FEAR efficiency to obtain large size (648 atom) ``\textit{ab initio}'' based models. FEAR allows system to evolve on it's own to find the appropriate energy minimum based on the force direction evaluated at each
relaxation step. This inclusion of \textit{ab initio} interactions not only guides us towards a chemically correct structures, it directly helps us to avoid
high energy small ring structures. A typical RMC based calculation fails to accurately model amorphous systems without the addition of experimental
based constraints. FEAR models yield a lower DFT energy minimum and take less time to converge as compared to the regular models obtained via. method of ``melt and quench'' with same interactions.

We have established a set of accurate \textit{ab initio} models for amorphous carbon that we hope will serve as a benchmark for future modeling studies.\\

\section{\label{Aknowledgment} Acknowledgment} 
The authors are thankful to the NSF under grant number DMR 1506836. We thank Dr. Ronald L. Cappelletti for helpful conversations. We acknowledge the financial support from 
Condensed Matter and Surface Science (CMSS) at Ohio University.
We also acknowledge computing time provided by the Ohio Supercomputer Center.

\section*{Data availability}
The coordinates of the four relaxed FEAR (648 atoms) models and of melt quench models prepared during this study are available from corresponding authors upon request. \\

\bibliography{sample}

\begin{thebibliography}{63}%
\makeatletter
\providecommand \@ifxundefined [1]{%
 \@ifx{#1\undefined}
}%
\providecommand \@ifnum [1]{%
 \ifnum #1\expandafter \@firstoftwo
 \else \expandafter \@secondoftwo
 \fi
}%
\providecommand \@ifx [1]{%
 \ifx #1\expandafter \@firstoftwo
 \else \expandafter \@secondoftwo
 \fi
}%
\providecommand \natexlab [1]{#1}%
\providecommand \enquote  [1]{``#1''}%
\providecommand \bibnamefont  [1]{#1}%
\providecommand \bibfnamefont [1]{#1}%
\providecommand \citenamefont [1]{#1}%
\providecommand \href@noop [0]{\@secondoftwo}%
\providecommand \href [0]{\begingroup \@sanitize@url \@href}%
\providecommand \@href[1]{\@@startlink{#1}\@@href}%
\providecommand \@@href[1]{\endgroup#1\@@endlink}%
\providecommand \@sanitize@url [0]{\catcode `\\12\catcode `\$12\catcode
  `\&12\catcode `\#12\catcode `\^12\catcode `\_12\catcode `\%12\relax}%
\providecommand \@@startlink[1]{}%
\providecommand \@@endlink[0]{}%
\providecommand \url  [0]{\begingroup\@sanitize@url \@url }%
\providecommand \@url [1]{\endgroup\@href {#1}{\urlprefix }}%
\providecommand \urlprefix  [0]{URL }%
\providecommand \Eprint [0]{\href }%
\providecommand \doibase [0]{http://dx.doi.org/}%
\providecommand \selectlanguage [0]{\@gobble}%
\providecommand \bibinfo  [0]{\@secondoftwo}%
\providecommand \bibfield  [0]{\@secondoftwo}%
\providecommand \translation [1]{[#1]}%
\providecommand \BibitemOpen [0]{}%
\providecommand \bibitemStop [0]{}%
\providecommand \bibitemNoStop [0]{.\EOS\space}%
\providecommand \EOS [0]{\spacefactor3000\relax}%
\providecommand \BibitemShut  [1]{\csname bibitem#1\endcsname}%
\let\auto@bib@innerbib\@empty
\bibitem [{\citenamefont {Mckenzie}(1995)}]{DRMcKenzie1}%
  \BibitemOpen
  \bibfield  {author} {\bibinfo {author} {\bibfnamefont {D.~R.}\ \bibnamefont
  {Mckenzie}},\ }\href@noop {} {\bibfield  {journal} {\bibinfo  {journal} {Rep.
  Prog. Phys.}\ }\textbf {\bibinfo {volume} {59}},\ \bibinfo {pages} {1611}
  (\bibinfo {year} {1995})}\BibitemShut {NoStop}%
\bibitem [{\citenamefont {Drabold}(2009)}]{DraboldEuro1}%
  \BibitemOpen
  \bibfield  {author} {\bibinfo {author} {\bibfnamefont {D.~A.}\ \bibnamefont
  {Drabold}},\ }\href@noop {} {\bibfield  {journal} {\bibinfo  {journal}
  {Eur.Phys.J. B}\ }\textbf {\bibinfo {volume} {68}},\ \bibinfo {pages} {1}
  (\bibinfo {year} {2009})}\BibitemShut {NoStop}%
\bibitem [{\citenamefont {Robertson}(2002)}]{Roberston1}%
  \BibitemOpen
  \bibfield  {author} {\bibinfo {author} {\bibfnamefont {J.}~\bibnamefont
  {Robertson}},\ }\href@noop {} {\bibfield  {journal} {\bibinfo  {journal}
  {Mater. Sci. Eng.}\ }\textbf {\bibinfo {volume} {R37}},\ \bibinfo {pages}
  {129} (\bibinfo {year} {2002})}\BibitemShut {NoStop}%
\bibitem [{\citenamefont {McGreevy}\ and\ \citenamefont
  {Pusztai}(1988)}]{MCGreevyPusztai1}%
  \BibitemOpen
  \bibfield  {author} {\bibinfo {author} {\bibfnamefont {R.~L.}\ \bibnamefont
  {McGreevy}}\ and\ \bibinfo {author} {\bibfnamefont {L.}~\bibnamefont
  {Pusztai}},\ }\href@noop {} {\bibfield  {journal} {\bibinfo  {journal} {Mol.
  Simul.}\ }\textbf {\bibinfo {volume} {1}},\ \bibinfo {pages} {359} (\bibinfo
  {year} {1988})}\BibitemShut {NoStop}%
\bibitem [{\citenamefont {Biswas}\ \emph {et~al.}(2004)\citenamefont {Biswas},
  \citenamefont {Atta-Fynn},\ and\ \citenamefont {Drabold}}]{Partha1}%
  \BibitemOpen
  \bibfield  {author} {\bibinfo {author} {\bibfnamefont {P.}~\bibnamefont
  {Biswas}}, \bibinfo {author} {\bibfnamefont {R.}~\bibnamefont {Atta-Fynn}}, \
  and\ \bibinfo {author} {\bibfnamefont {D.~A.}\ \bibnamefont {Drabold}},\
  }\href@noop {} {\bibfield  {journal} {\bibinfo  {journal} {Phys.Rev.B}\
  }\textbf {\bibinfo {volume} {69}},\ \bibinfo {pages} {195207} (\bibinfo
  {year} {2004})}\BibitemShut {NoStop}%
\bibitem [{\citenamefont {O'Malley}\ \emph {et~al.}(1998)\citenamefont
  {O'Malley}, \citenamefont {Snook},\ and\ \citenamefont
  {McCulloch}}]{Malley1}%
  \BibitemOpen
  \bibfield  {author} {\bibinfo {author} {\bibfnamefont {B.}~\bibnamefont
  {O'Malley}}, \bibinfo {author} {\bibfnamefont {I.}~\bibnamefont {Snook}}, \
  and\ \bibinfo {author} {\bibfnamefont {D.}~\bibnamefont {McCulloch}},\
  }\href@noop {} {\bibfield  {journal} {\bibinfo  {journal} {Phys.Rev.B}\
  }\textbf {\bibinfo {volume} {57}},\ \bibinfo {pages} {14148} (\bibinfo {year}
  {1998})}\BibitemShut {NoStop}%
\bibitem [{\citenamefont {Keen}\ and\ \citenamefont {Dove}(1999)}]{Keen1}%
  \BibitemOpen
  \bibfield  {author} {\bibinfo {author} {\bibfnamefont {D.~A.}\ \bibnamefont
  {Keen}}\ and\ \bibinfo {author} {\bibfnamefont {M.~T.}\ \bibnamefont
  {Dove}},\ }\href@noop {} {\bibfield  {journal} {\bibinfo  {journal} {J.
  Phys.: Condens. Matter}\ }\textbf {\bibinfo {volume} {11}},\ \bibinfo {pages}
  {9263} (\bibinfo {year} {1999})}\BibitemShut {NoStop}%
\bibitem [{\citenamefont {Gereben}\ and\ \citenamefont
  {Pusztai}(1994)}]{GerebenPusztai1}%
  \BibitemOpen
  \bibfield  {author} {\bibinfo {author} {\bibfnamefont {V.}~\bibnamefont
  {Gereben}}\ and\ \bibinfo {author} {\bibfnamefont {L.}~\bibnamefont
  {Pusztai}},\ }\href@noop {} {\bibfield  {journal} {\bibinfo  {journal}
  {Phys.Rev.B}\ }\textbf {\bibinfo {volume} {50}},\ \bibinfo {pages} {14136}
  (\bibinfo {year} {1994})}\BibitemShut {NoStop}%
\bibitem [{\citenamefont {Opletal}\ \emph {et~al.}(2017)\citenamefont
  {Opletal}, \citenamefont {Petersen}, \citenamefont {Barnard},\ and\
  \citenamefont {Russo}}]{Opletal2}%
  \BibitemOpen
  \bibfield  {author} {\bibinfo {author} {\bibfnamefont {G.}~\bibnamefont
  {Opletal}}, \bibinfo {author} {\bibfnamefont {T.~C.}\ \bibnamefont
  {Petersen}}, \bibinfo {author} {\bibfnamefont {A.~S.}\ \bibnamefont
  {Barnard}}, \ and\ \bibinfo {author} {\bibfnamefont {S.~P.}\ \bibnamefont
  {Russo}},\ }\href@noop {} {\bibfield  {journal} {\bibinfo  {journal} {J.
  Comput. Chem.}\ }\textbf {\bibinfo {volume} {38}},\ \bibinfo {pages} {1547}
  (\bibinfo {year} {2017})}\BibitemShut {NoStop}%
\bibitem [{\citenamefont {Jain}\ \emph {et~al.}(2006)\citenamefont {Jain},
  \citenamefont {Pellenq}, \citenamefont {Pikunic},\ and\ \citenamefont
  {Gubbins}}]{Jain1}%
  \BibitemOpen
  \bibfield  {author} {\bibinfo {author} {\bibfnamefont {S.~K.}\ \bibnamefont
  {Jain}}, \bibinfo {author} {\bibfnamefont {R.~J.~M.}\ \bibnamefont
  {Pellenq}}, \bibinfo {author} {\bibfnamefont {J.~P.}\ \bibnamefont
  {Pikunic}}, \ and\ \bibinfo {author} {\bibfnamefont {K.~E.}\ \bibnamefont
  {Gubbins}},\ }\href@noop {} {\bibfield  {journal} {\bibinfo  {journal}
  {Langmuir}\ }\textbf {\bibinfo {volume} {22}},\ \bibinfo {pages} {9942}
  (\bibinfo {year} {2006})}\BibitemShut {NoStop}%
\bibitem [{\citenamefont {Walters}\ \emph {et~al.}(1997)\citenamefont
  {Walters}, \citenamefont {Gilkes}, \citenamefont {Wicks},\ and\ \citenamefont
  {Newport}}]{JKWalters1}%
  \BibitemOpen
  \bibfield  {author} {\bibinfo {author} {\bibfnamefont {J.~K.}\ \bibnamefont
  {Walters}}, \bibinfo {author} {\bibfnamefont {K.~W.~R.}\ \bibnamefont
  {Gilkes}}, \bibinfo {author} {\bibfnamefont {J.~D.}\ \bibnamefont {Wicks}}, \
  and\ \bibinfo {author} {\bibfnamefont {R.~J.}\ \bibnamefont {Newport}},\
  }\href@noop {} {\bibfield  {journal} {\bibinfo  {journal} {J. Phys.: Condens.
  Matter}\ }\textbf {\bibinfo {volume} {9}},\ \bibinfo {pages} {457} (\bibinfo
  {year} {1997})}\BibitemShut {NoStop}%
\bibitem [{\citenamefont {Hosokawa}\ \emph {et~al.}(2012)\citenamefont
  {Hosokawa}, \citenamefont {Pilgrim}, \citenamefont {Berar},\ and\
  \citenamefont {Kohara}}]{Hosokawa1}%
  \BibitemOpen
  \bibfield  {author} {\bibinfo {author} {\bibfnamefont {S.}~\bibnamefont
  {Hosokawa}}, \bibinfo {author} {\bibfnamefont {W.~C.}\ \bibnamefont
  {Pilgrim}}, \bibinfo {author} {\bibfnamefont {J.~F.}\ \bibnamefont {Berar}},
  \ and\ \bibinfo {author} {\bibfnamefont {S.}~\bibnamefont {Kohara}},\
  }\href@noop {} {\bibfield  {journal} {\bibinfo  {journal} {Eur. Phys. J.
  Spec. Top.}\ }\textbf {\bibinfo {volume} {208}},\ \bibinfo {pages} {291}
  (\bibinfo {year} {2012})}\BibitemShut {NoStop}%
\bibitem [{\citenamefont {Gurman}\ and\ \citenamefont
  {McGreevy}(1990)}]{Gurman1}%
  \BibitemOpen
  \bibfield  {author} {\bibinfo {author} {\bibfnamefont {S.~J.}\ \bibnamefont
  {Gurman}}\ and\ \bibinfo {author} {\bibfnamefont {R.~L.}\ \bibnamefont
  {McGreevy}},\ }\href@noop {} {\bibfield  {journal} {\bibinfo  {journal} {J.
  Phys.: Condens. Matter}\ }\textbf {\bibinfo {volume} {2}},\ \bibinfo {pages}
  {9463} (\bibinfo {year} {1990})}\BibitemShut {NoStop}%
\bibitem [{\citenamefont {Tucker}\ \emph {et~al.}(2005)\citenamefont {Tucker},
  \citenamefont {Keen}, \citenamefont {Dove},\ and\ \citenamefont
  {Trachenko}}]{Tucker1}%
  \BibitemOpen
  \bibfield  {author} {\bibinfo {author} {\bibfnamefont {M.~G.}\ \bibnamefont
  {Tucker}}, \bibinfo {author} {\bibfnamefont {D.~A.}\ \bibnamefont {Keen}},
  \bibinfo {author} {\bibfnamefont {M.~T.}\ \bibnamefont {Dove}}, \ and\
  \bibinfo {author} {\bibfnamefont {K.}~\bibnamefont {Trachenko}},\ }\href@noop
  {} {\bibfield  {journal} {\bibinfo  {journal} {J. Phys.: Condens. Matter}\
  }\textbf {\bibinfo {volume} {17}},\ \bibinfo {pages} {67} (\bibinfo {year}
  {2005})}\BibitemShut {NoStop}%
\bibitem [{\citenamefont {Biswas}\ \emph {et~al.}(2005)\citenamefont {Biswas},
  \citenamefont {Tafen},\ and\ \citenamefont {Drabold}}]{Partha2}%
  \BibitemOpen
  \bibfield  {author} {\bibinfo {author} {\bibfnamefont {P.}~\bibnamefont
  {Biswas}}, \bibinfo {author} {\bibfnamefont {D.~N.}\ \bibnamefont {Tafen}}, \
  and\ \bibinfo {author} {\bibfnamefont {D.~A.}\ \bibnamefont {Drabold}},\
  }\href@noop {} {\bibfield  {journal} {\bibinfo  {journal} {Phys.Rev.B}\
  }\textbf {\bibinfo {volume} {71}},\ \bibinfo {pages} {054204} (\bibinfo
  {year} {2005})}\BibitemShut {NoStop}%
\bibitem [{\citenamefont {Opletal}\ \emph {et~al.}(2002)\citenamefont
  {Opletal}, \citenamefont {Petersen}, \citenamefont {Omalley}, \citenamefont
  {Snook}, \citenamefont {Mcculloch}, \citenamefont {Marks},\ and\
  \citenamefont {Yarovsky}}]{Opletal5}%
  \BibitemOpen
  \bibfield  {author} {\bibinfo {author} {\bibfnamefont {G.}~\bibnamefont
  {Opletal}}, \bibinfo {author} {\bibfnamefont {T.}~\bibnamefont {Petersen}},
  \bibinfo {author} {\bibfnamefont {B.}~\bibnamefont {Omalley}}, \bibinfo
  {author} {\bibfnamefont {I.}~\bibnamefont {Snook}}, \bibinfo {author}
  {\bibfnamefont {D.~G.}\ \bibnamefont {Mcculloch}}, \bibinfo {author}
  {\bibfnamefont {N.~A.}\ \bibnamefont {Marks}}, \ and\ \bibinfo {author}
  {\bibfnamefont {I.}~\bibnamefont {Yarovsky}},\ }\href@noop {} {\bibfield
  {journal} {\bibinfo  {journal} {Mol. Sim.}\ }\textbf {\bibinfo {volume}
  {28}},\ \bibinfo {pages} {927} (\bibinfo {year} {2002})}\BibitemShut
  {NoStop}%
\bibitem [{\citenamefont {Pandey}\ \emph
  {et~al.}(2016{\natexlab{a}})\citenamefont {Pandey}, \citenamefont {Biswas},\
  and\ \citenamefont {Drabold}}]{Anup1}%
  \BibitemOpen
  \bibfield  {author} {\bibinfo {author} {\bibfnamefont {A.}~\bibnamefont
  {Pandey}}, \bibinfo {author} {\bibfnamefont {P.}~\bibnamefont {Biswas}}, \
  and\ \bibinfo {author} {\bibfnamefont {D.~A.}\ \bibnamefont {Drabold}},\
  }\href@noop {} {\bibfield  {journal} {\bibinfo  {journal} {Scientific
  Reports}\ }\textbf {\bibinfo {volume} {6}},\ \bibinfo {pages} {33731}
  (\bibinfo {year} {2016}{\natexlab{a}})}\BibitemShut {NoStop}%
\bibitem [{\citenamefont {Pandey}\ \emph {et~al.}(2015)\citenamefont {Pandey},
  \citenamefont {Biswas},\ and\ \citenamefont {Drabold}}]{Anup2}%
  \BibitemOpen
  \bibfield  {author} {\bibinfo {author} {\bibfnamefont {A.}~\bibnamefont
  {Pandey}}, \bibinfo {author} {\bibfnamefont {P.}~\bibnamefont {Biswas}}, \
  and\ \bibinfo {author} {\bibfnamefont {D.~A.}\ \bibnamefont {Drabold}},\
  }\href@noop {} {\bibfield  {journal} {\bibinfo  {journal} {Phys.Rev.B}\
  }\textbf {\bibinfo {volume} {92}},\ \bibinfo {pages} {155205} (\bibinfo
  {year} {2015})}\BibitemShut {NoStop}%
\bibitem [{\citenamefont {Pandey}\ \emph
  {et~al.}(2016{\natexlab{b}})\citenamefont {Pandey}, \citenamefont {Biswas},
  \citenamefont {Bhattarai},\ and\ \citenamefont {Drabold}}]{Anup3}%
  \BibitemOpen
  \bibfield  {author} {\bibinfo {author} {\bibfnamefont {A.}~\bibnamefont
  {Pandey}}, \bibinfo {author} {\bibfnamefont {P.}~\bibnamefont {Biswas}},
  \bibinfo {author} {\bibfnamefont {B.}~\bibnamefont {Bhattarai}}, \ and\
  \bibinfo {author} {\bibfnamefont {D.~A.}\ \bibnamefont {Drabold}},\
  }\href@noop {} {\bibfield  {journal} {\bibinfo  {journal} {Phys.Rev.B}\
  }\textbf {\bibinfo {volume} {94}},\ \bibinfo {pages} {235208} (\bibinfo
  {year} {2016}{\natexlab{b}})}\BibitemShut {NoStop}%
\bibitem [{\citenamefont {Prasai}\ \emph {et~al.}(2016)\citenamefont {Prasai},
  \citenamefont {Biswas},\ and\ \citenamefont {Drabold}}]{Kiran1}%
  \BibitemOpen
  \bibfield  {author} {\bibinfo {author} {\bibfnamefont {K.}~\bibnamefont
  {Prasai}}, \bibinfo {author} {\bibfnamefont {P.}~\bibnamefont {Biswas}}, \
  and\ \bibinfo {author} {\bibfnamefont {D.~A.}\ \bibnamefont {Drabold}},\
  }\href@noop {} {\bibfield  {journal} {\bibinfo  {journal} {Phys. Status
  Solidi A}\ }\textbf {\bibinfo {volume} {213}},\ \bibinfo {pages} {1653}
  (\bibinfo {year} {2016})}\BibitemShut {NoStop}%
\bibitem [{\citenamefont {Cliffe}\ \emph {et~al.}(2017)\citenamefont {Cliffe},
  \citenamefont {Bartok}, \citenamefont {Kerber}, \citenamefont {Grey},
  \citenamefont {Csanyi},\ and\ \citenamefont {Goodwin}}]{Cliffe1}%
  \BibitemOpen
  \bibfield  {author} {\bibinfo {author} {\bibfnamefont {M.~J.}\ \bibnamefont
  {Cliffe}}, \bibinfo {author} {\bibfnamefont {A.~P.}\ \bibnamefont {Bartok}},
  \bibinfo {author} {\bibfnamefont {R.~N.}\ \bibnamefont {Kerber}}, \bibinfo
  {author} {\bibfnamefont {C.~P.}\ \bibnamefont {Grey}}, \bibinfo {author}
  {\bibfnamefont {G.}~\bibnamefont {Csanyi}}, \ and\ \bibinfo {author}
  {\bibfnamefont {A.~L.}\ \bibnamefont {Goodwin}},\ }\href@noop {} {\bibfield
  {journal} {\bibinfo  {journal} {Phys.Rev.B}\ }\textbf {\bibinfo {volume}
  {95}},\ \bibinfo {pages} {224108} (\bibinfo {year} {2017})}\BibitemShut
  {NoStop}%
\bibitem [{\citenamefont {Cliffe}\ \emph {et~al.}(2010)\citenamefont {Cliffe},
  \citenamefont {Dove}, \citenamefont {Drabold},\ and\ \citenamefont
  {Goodwin}}]{Cliffe2}%
  \BibitemOpen
  \bibfield  {author} {\bibinfo {author} {\bibfnamefont {M.~J.}\ \bibnamefont
  {Cliffe}}, \bibinfo {author} {\bibfnamefont {M.~T.}\ \bibnamefont {Dove}},
  \bibinfo {author} {\bibfnamefont {D.~A.}\ \bibnamefont {Drabold}}, \ and\
  \bibinfo {author} {\bibfnamefont {A.~L.}\ \bibnamefont {Goodwin}},\
  }\href@noop {} {\bibfield  {journal} {\bibinfo  {journal} {Phys.Rev.Lett}\
  }\textbf {\bibinfo {volume} {104}},\ \bibinfo {pages} {125501} (\bibinfo
  {year} {2010})}\BibitemShut {NoStop}%
\bibitem [{\citenamefont {Tersoff}(1988)}]{Tersoff1}%
  \BibitemOpen
  \bibfield  {author} {\bibinfo {author} {\bibfnamefont {J.}~\bibnamefont
  {Tersoff}},\ }\href@noop {} {\bibfield  {journal} {\bibinfo  {journal} {Phys.
  Rev. Lett.}\ }\textbf {\bibinfo {volume} {61}},\ \bibinfo {pages} {2879}
  (\bibinfo {year} {1988})}\BibitemShut {NoStop}%
\bibitem [{\citenamefont {Marks}(2000)}]{NAMarks1}%
  \BibitemOpen
  \bibfield  {author} {\bibinfo {author} {\bibfnamefont {N.~A.}\ \bibnamefont
  {Marks}},\ }\href@noop {} {\bibfield  {journal} {\bibinfo  {journal}
  {Phys.Rev.B}\ }\textbf {\bibinfo {volume} {63}},\ \bibinfo {pages} {035401}
  (\bibinfo {year} {2000})}\BibitemShut {NoStop}%
\bibitem [{\citenamefont {Mathioudakis}\ \emph {et~al.}(2004)\citenamefont
  {Mathioudakis}, \citenamefont {Kopidakis}, \citenamefont {Kelires},
  \citenamefont {Wang},\ and\ \citenamefont {Ho}}]{TBMD2}%
  \BibitemOpen
  \bibfield  {author} {\bibinfo {author} {\bibfnamefont {C.}~\bibnamefont
  {Mathioudakis}}, \bibinfo {author} {\bibfnamefont {G.}~\bibnamefont
  {Kopidakis}}, \bibinfo {author} {\bibfnamefont {P.~C.}\ \bibnamefont
  {Kelires}}, \bibinfo {author} {\bibfnamefont {C.~Z.}\ \bibnamefont {Wang}}, \
  and\ \bibinfo {author} {\bibfnamefont {K.~M.}\ \bibnamefont {Ho}},\
  }\href@noop {} {\bibfield  {journal} {\bibinfo  {journal} {Phys.Rev.B}\
  }\textbf {\bibinfo {volume} {70}},\ \bibinfo {pages} {125202} (\bibinfo
  {year} {2004})}\BibitemShut {NoStop}%
\bibitem [{\citenamefont {Li}\ \emph {et~al.}(2013)\citenamefont {Li},
  \citenamefont {Xu}, \citenamefont {Song}, \citenamefont {Ovcharenko},
  \citenamefont {Zhang},\ and\ \citenamefont {Jia}}]{LLi1}%
  \BibitemOpen
  \bibfield  {author} {\bibinfo {author} {\bibfnamefont {L.}~\bibnamefont
  {Li}}, \bibinfo {author} {\bibfnamefont {M.}~\bibnamefont {Xu}}, \bibinfo
  {author} {\bibfnamefont {W.}~\bibnamefont {Song}}, \bibinfo {author}
  {\bibfnamefont {A.}~\bibnamefont {Ovcharenko}}, \bibinfo {author}
  {\bibfnamefont {G.}~\bibnamefont {Zhang}}, \ and\ \bibinfo {author}
  {\bibfnamefont {D.}~\bibnamefont {Jia}},\ }\href@noop {} {\bibfield
  {journal} {\bibinfo  {journal} {App. Surf. Sci.}\ }\textbf {\bibinfo {volume}
  {286}},\ \bibinfo {pages} {287} (\bibinfo {year} {2013})}\BibitemShut
  {NoStop}%
\bibitem [{Note1()}]{Note1}%
  \BibitemOpen
  \bibinfo {note} {DFT code using LDA with Ceperley Alder exchange
  correlation\cite {siesta}}\BibitemShut {NoStop}%
\bibitem [{\citenamefont {Kresse}\ and\ \citenamefont
  {Furthmuller}(1996)}]{Kresse2}%
  \BibitemOpen
  \bibfield  {author} {\bibinfo {author} {\bibfnamefont {G.}~\bibnamefont
  {Kresse}}\ and\ \bibinfo {author} {\bibfnamefont {J.}~\bibnamefont
  {Furthmuller}},\ }\href@noop {} {\bibfield  {journal} {\bibinfo  {journal}
  {Phys.Rev.B}\ }\textbf {\bibinfo {volume} {54}},\ \bibinfo {pages} {11169}
  (\bibinfo {year} {1996})}\BibitemShut {NoStop}%
\bibitem [{\citenamefont {Kresse}\ and\ \citenamefont
  {Joubert}(1999)}]{Kresse1}%
  \BibitemOpen
  \bibfield  {author} {\bibinfo {author} {\bibfnamefont {G.}~\bibnamefont
  {Kresse}}\ and\ \bibinfo {author} {\bibfnamefont {D.}~\bibnamefont
  {Joubert}},\ }\href@noop {} {\bibfield  {journal} {\bibinfo  {journal}
  {Phys.Rev.B}\ }\textbf {\bibinfo {volume} {59}},\ \bibinfo {pages} {1758}
  (\bibinfo {year} {1999})}\BibitemShut {NoStop}%
\bibitem [{\citenamefont {Blochl}(1994)}]{Bloch1}%
  \BibitemOpen
  \bibfield  {author} {\bibinfo {author} {\bibfnamefont {P.~E.}\ \bibnamefont
  {Blochl}},\ }\href@noop {} {\bibfield  {journal} {\bibinfo  {journal}
  {Phys.Rev.B}\ }\textbf {\bibinfo {volume} {50}},\ \bibinfo {pages} {17953}
  (\bibinfo {year} {1994})}\BibitemShut {NoStop}%
\bibitem [{Note2()}]{Note2}%
  \BibitemOpen
  \bibinfo {note} {RMC based applications for the structural refinement\cite
  {Tucker2}}\BibitemShut {NoStop}%
\bibitem [{\citenamefont {Bhattarai}\ and\ \citenamefont
  {Drabold}(2017)}]{Bhattarai2}%
  \BibitemOpen
  \bibfield  {author} {\bibinfo {author} {\bibfnamefont {B.}~\bibnamefont
  {Bhattarai}}\ and\ \bibinfo {author} {\bibfnamefont {D.~A.}\ \bibnamefont
  {Drabold}},\ }\href@noop {} {\bibfield  {journal} {\bibinfo  {journal}
  {Carbon}\ }\textbf {\bibinfo {volume} {115}},\ \bibinfo {pages} {532}
  (\bibinfo {year} {2017})}\BibitemShut {NoStop}%
\bibitem [{\citenamefont {Opletal}\ \emph {et~al.}(2005)\citenamefont
  {Opletal}, \citenamefont {Petersen}, \citenamefont {McCulloch}, \citenamefont
  {Snook},\ and\ \citenamefont {Yarovsky}}]{Opletal1}%
  \BibitemOpen
  \bibfield  {author} {\bibinfo {author} {\bibfnamefont {G.}~\bibnamefont
  {Opletal}}, \bibinfo {author} {\bibfnamefont {T.~C.}\ \bibnamefont
  {Petersen}}, \bibinfo {author} {\bibfnamefont {D.~G.}\ \bibnamefont
  {McCulloch}}, \bibinfo {author} {\bibfnamefont {I.~K.}\ \bibnamefont
  {Snook}}, \ and\ \bibinfo {author} {\bibfnamefont {I.}~\bibnamefont
  {Yarovsky}},\ }\href@noop {} {\bibfield  {journal} {\bibinfo  {journal} {J.
  Phys: Condens. Matter}\ }\textbf {\bibinfo {volume} {17}},\ \bibinfo {pages}
  {2605} (\bibinfo {year} {2005})}\BibitemShut {NoStop}%
\bibitem [{\citenamefont {Djordjevic}\ \emph {et~al.}(1995)\citenamefont
  {Djordjevic}, \citenamefont {Thorpe},\ and\ \citenamefont {Wooten}}]{WWW1}%
  \BibitemOpen
  \bibfield  {author} {\bibinfo {author} {\bibfnamefont {B.}~\bibnamefont
  {Djordjevic}}, \bibinfo {author} {\bibfnamefont {M.}~\bibnamefont {Thorpe}},
  \ and\ \bibinfo {author} {\bibfnamefont {F.}~\bibnamefont {Wooten}},\
  }\href@noop {} {\bibfield  {journal} {\bibinfo  {journal} {Phys.Rev.B}\
  }\textbf {\bibinfo {volume} {52}},\ \bibinfo {pages} {5685} (\bibinfo {year}
  {1995})}\BibitemShut {NoStop}%
\bibitem [{\citenamefont {Gilkes}\ \emph {et~al.}(1995)\citenamefont {Gilkes},
  \citenamefont {Gaskell},\ and\ \citenamefont {Robertson}}]{Gilkes1}%
  \BibitemOpen
  \bibfield  {author} {\bibinfo {author} {\bibfnamefont {K.~W.~R.}\
  \bibnamefont {Gilkes}}, \bibinfo {author} {\bibfnamefont {P.~H.}\
  \bibnamefont {Gaskell}}, \ and\ \bibinfo {author} {\bibfnamefont
  {J.}~\bibnamefont {Robertson}},\ }\href@noop {} {\bibfield  {journal}
  {\bibinfo  {journal} {Phys.Rev.B}\ }\textbf {\bibinfo {volume} {51}},\
  \bibinfo {pages} {12303} (\bibinfo {year} {1995})}\BibitemShut {NoStop}%
\bibitem [{\citenamefont {Li}\ and\ \citenamefont {Lannin}(1990)}]{FangLi1}%
  \BibitemOpen
  \bibfield  {author} {\bibinfo {author} {\bibfnamefont {F.}~\bibnamefont
  {Li}}\ and\ \bibinfo {author} {\bibfnamefont {J.}~\bibnamefont {Lannin}},\
  }\href@noop {} {\bibfield  {journal} {\bibinfo  {journal} {Phys.Rev.Lett}\
  }\textbf {\bibinfo {volume} {65}},\ \bibinfo {pages} {1905} (\bibinfo {year}
  {1990})}\BibitemShut {NoStop}%
\bibitem [{\citenamefont {Farmahini}\ \emph {et~al.}(2013)\citenamefont
  {Farmahini}, \citenamefont {Opletal},\ and\ \citenamefont
  {Bhatia}}]{Opletal4}%
  \BibitemOpen
  \bibfield  {author} {\bibinfo {author} {\bibfnamefont {A.~H.}\ \bibnamefont
  {Farmahini}}, \bibinfo {author} {\bibfnamefont {G.}~\bibnamefont {Opletal}},
  \ and\ \bibinfo {author} {\bibfnamefont {S.~K.}\ \bibnamefont {Bhatia}},\
  }\href@noop {} {\bibfield  {journal} {\bibinfo  {journal} {J. Phys. Chem. C}\
  }\textbf {\bibinfo {volume} {117}},\ \bibinfo {pages} {14081−14094}
  (\bibinfo {year} {2013})}\BibitemShut {NoStop}%
\bibitem [{\citenamefont {McCulloch}\ \emph {et~al.}(2000)\citenamefont
  {McCulloch}, \citenamefont {McKenzie},\ and\ \citenamefont
  {Goringe}}]{McCulloch1}%
  \BibitemOpen
  \bibfield  {author} {\bibinfo {author} {\bibfnamefont {D.~G.}\ \bibnamefont
  {McCulloch}}, \bibinfo {author} {\bibfnamefont {D.}~\bibnamefont {McKenzie}},
  \ and\ \bibinfo {author} {\bibfnamefont {C.}~\bibnamefont {Goringe}},\
  }\href@noop {} {\bibfield  {journal} {\bibinfo  {journal} {Phys.Rev.B}\
  }\textbf {\bibinfo {volume} {61}},\ \bibinfo {pages} {2349} (\bibinfo {year}
  {2000})}\BibitemShut {NoStop}%
\bibitem [{Note3()}]{Note3}%
  \BibitemOpen
  \bibinfo {note} {Jmol, an open-source Java viewer for chemical structures in
  3D}\BibitemShut {NoStop}%
\bibitem [{\citenamefont {Ferrari}\ \emph {et~al.}(2000)\citenamefont
  {Ferrari}, \citenamefont {Libassi}, \citenamefont {Tanner}, \citenamefont
  {Stolojan}, \citenamefont {Yuan}, \citenamefont {Brown}, \citenamefont
  {Rodil}, \citenamefont {Kleinsorge},\ and\ \citenamefont
  {Robertson}}]{ACFerrari1}%
  \BibitemOpen
  \bibfield  {author} {\bibinfo {author} {\bibfnamefont {A.~C.}\ \bibnamefont
  {Ferrari}}, \bibinfo {author} {\bibfnamefont {A.}~\bibnamefont {Libassi}},
  \bibinfo {author} {\bibfnamefont {B.~K.}\ \bibnamefont {Tanner}}, \bibinfo
  {author} {\bibfnamefont {V.}~\bibnamefont {Stolojan}}, \bibinfo {author}
  {\bibfnamefont {J.}~\bibnamefont {Yuan}}, \bibinfo {author} {\bibfnamefont
  {L.~M.}\ \bibnamefont {Brown}}, \bibinfo {author} {\bibfnamefont {S.~E.}\
  \bibnamefont {Rodil}}, \bibinfo {author} {\bibfnamefont {B.}~\bibnamefont
  {Kleinsorge}}, \ and\ \bibinfo {author} {\bibfnamefont {J.}~\bibnamefont
  {Robertson}},\ }\href@noop {} {\bibfield  {journal} {\bibinfo  {journal}
  {Phys.Rev.B}\ }\textbf {\bibinfo {volume} {62}},\ \bibinfo {pages} {11089}
  (\bibinfo {year} {2000})}\BibitemShut {NoStop}%
\bibitem [{\citenamefont {Fallon}\ \emph {et~al.}(1993)\citenamefont {Fallon},
  \citenamefont {Veerasamy}, \citenamefont {Davis}, \citenamefont {Robertson},
  \citenamefont {Amaratunga}, \citenamefont {Milne},\ and\ \citenamefont
  {Koskinen}}]{PJFAllon1}%
  \BibitemOpen
  \bibfield  {author} {\bibinfo {author} {\bibfnamefont {P.~J.}\ \bibnamefont
  {Fallon}}, \bibinfo {author} {\bibfnamefont {V.~S.}\ \bibnamefont
  {Veerasamy}}, \bibinfo {author} {\bibfnamefont {C.~A.}\ \bibnamefont
  {Davis}}, \bibinfo {author} {\bibfnamefont {J.}~\bibnamefont {Robertson}},
  \bibinfo {author} {\bibfnamefont {G.~A.~J.}\ \bibnamefont {Amaratunga}},
  \bibinfo {author} {\bibfnamefont {W.~I.}\ \bibnamefont {Milne}}, \ and\
  \bibinfo {author} {\bibfnamefont {J.}~\bibnamefont {Koskinen}},\ }\href@noop
  {} {\bibfield  {journal} {\bibinfo  {journal} {Phys.Rev.B}\ }\textbf
  {\bibinfo {volume} {48}},\ \bibinfo {pages} {4777} (\bibinfo {year}
  {1993})}\BibitemShut {NoStop}%
\bibitem [{\citenamefont {Wooten}\ \emph {et~al.}(1985)\citenamefont {Wooten},
  \citenamefont {Winer},\ and\ \citenamefont {Weaire}}]{WWW2}%
  \BibitemOpen
  \bibfield  {author} {\bibinfo {author} {\bibfnamefont {F.}~\bibnamefont
  {Wooten}}, \bibinfo {author} {\bibfnamefont {K.}~\bibnamefont {Winer}}, \
  and\ \bibinfo {author} {\bibfnamefont {D.}~\bibnamefont {Weaire}},\
  }\href@noop {} {\bibfield  {journal} {\bibinfo  {journal} {Phys.Rev.Lett.}\
  }\textbf {\bibinfo {volume} {54}},\ \bibinfo {pages} {1392} (\bibinfo {year}
  {1985})}\BibitemShut {NoStop}%
\bibitem [{\citenamefont {Tafen}\ and\ \citenamefont
  {Drabold}(2003)}]{TafenC1}%
  \BibitemOpen
  \bibfield  {author} {\bibinfo {author} {\bibfnamefont {D.~N.}\ \bibnamefont
  {Tafen}}\ and\ \bibinfo {author} {\bibfnamefont {D.~A.}\ \bibnamefont
  {Drabold}},\ }\href@noop {} {\bibfield  {journal} {\bibinfo  {journal}
  {Phys.Rev.B}\ }\textbf {\bibinfo {volume} {68}},\ \bibinfo {pages} {165208}
  (\bibinfo {year} {2003})}\BibitemShut {NoStop}%
\bibitem [{\citenamefont {Wang}\ and\ \citenamefont {Ho}(1994)}]{Wang2}%
  \BibitemOpen
  \bibfield  {author} {\bibinfo {author} {\bibfnamefont {C.}~\bibnamefont
  {Wang}}\ and\ \bibinfo {author} {\bibfnamefont {K.}~\bibnamefont {Ho}},\
  }\href@noop {} {\bibfield  {journal} {\bibinfo  {journal} {Phys.Rev.B}\
  }\textbf {\bibinfo {volume} {50}},\ \bibinfo {pages} {12429} (\bibinfo {year}
  {1994})}\BibitemShut {NoStop}%
\bibitem [{\citenamefont {T.Petersen}\ \emph {et~al.}(2003)\citenamefont
  {T.Petersen}, \citenamefont {Yarovsky}, \citenamefont {Snook}, \citenamefont
  {McCulloch},\ and\ \citenamefont {Opletal}}]{Opletal3}%
  \BibitemOpen
  \bibfield  {author} {\bibinfo {author} {\bibnamefont {T.Petersen}}, \bibinfo
  {author} {\bibfnamefont {I.}~\bibnamefont {Yarovsky}}, \bibinfo {author}
  {\bibfnamefont {I.}~\bibnamefont {Snook}}, \bibinfo {author} {\bibfnamefont
  {D.~G.}\ \bibnamefont {McCulloch}}, \ and\ \bibinfo {author} {\bibfnamefont
  {G.}~\bibnamefont {Opletal}},\ }\href@noop {} {\bibfield  {journal} {\bibinfo
   {journal} {Carbon}\ }\textbf {\bibinfo {volume} {41}},\ \bibinfo {pages}
  {2403} (\bibinfo {year} {2003})}\BibitemShut {NoStop}%
\bibitem [{\citenamefont {Mildner}\ and\ \citenamefont
  {Carpenter}(1982)}]{Mildner1}%
  \BibitemOpen
  \bibfield  {author} {\bibinfo {author} {\bibfnamefont {D.}~\bibnamefont
  {Mildner}}\ and\ \bibinfo {author} {\bibfnamefont {J.}~\bibnamefont
  {Carpenter}},\ }\href@noop {} {\bibfield  {journal} {\bibinfo  {journal}
  {Journal of Non-Crystalline Solids}\ }\textbf {\bibinfo {volume} {47}},\
  \bibinfo {pages} {391} (\bibinfo {year} {1982})}\BibitemShut {NoStop}%
\bibitem [{\citenamefont {Zetterstrom}\ \emph {et~al.}(2005)\citenamefont
  {Zetterstrom}, \citenamefont {Urbonaite}, \citenamefont {Lindberg},
  \citenamefont {Delaplane}, \citenamefont {Leis},\ and\ \citenamefont
  {Svensson}}]{Zetterstrom1}%
  \BibitemOpen
  \bibfield  {author} {\bibinfo {author} {\bibfnamefont {P.}~\bibnamefont
  {Zetterstrom}}, \bibinfo {author} {\bibfnamefont {S.}~\bibnamefont
  {Urbonaite}}, \bibinfo {author} {\bibfnamefont {F.}~\bibnamefont {Lindberg}},
  \bibinfo {author} {\bibfnamefont {R.~G.}\ \bibnamefont {Delaplane}}, \bibinfo
  {author} {\bibfnamefont {J.}~\bibnamefont {Leis}}, \ and\ \bibinfo {author}
  {\bibfnamefont {G.}~\bibnamefont {Svensson}},\ }\href@noop {} {\bibfield
  {journal} {\bibinfo  {journal} {J. Phys.: Condens. Matter}\ }\textbf
  {\bibinfo {volume} {17}},\ \bibinfo {pages} {3509} (\bibinfo {year}
  {2005})}\BibitemShut {NoStop}%
\bibitem [{\citenamefont {Pikunic}\ \emph {et~al.}(2003)\citenamefont
  {Pikunic}, \citenamefont {Clinard}, \citenamefont {Cohaut}, \citenamefont
  {Gubbins}, \citenamefont {Guet}, \citenamefont {Pellenq}, \citenamefont
  {Rannou},\ and\ \citenamefont {Rouzaud}}]{Pikunic1}%
  \BibitemOpen
  \bibfield  {author} {\bibinfo {author} {\bibfnamefont {J.}~\bibnamefont
  {Pikunic}}, \bibinfo {author} {\bibfnamefont {C.}~\bibnamefont {Clinard}},
  \bibinfo {author} {\bibfnamefont {N.}~\bibnamefont {Cohaut}}, \bibinfo
  {author} {\bibfnamefont {K.~E.}\ \bibnamefont {Gubbins}}, \bibinfo {author}
  {\bibfnamefont {J.-M.}\ \bibnamefont {Guet}}, \bibinfo {author}
  {\bibfnamefont {R.-M.}\ \bibnamefont {Pellenq}}, \bibinfo {author}
  {\bibfnamefont {I.}~\bibnamefont {Rannou}}, \ and\ \bibinfo {author}
  {\bibfnamefont {J.-N.}\ \bibnamefont {Rouzaud}},\ }\href@noop {} {\bibfield
  {journal} {\bibinfo  {journal} {Langmuir}\ }\textbf {\bibinfo {volume}
  {19}},\ \bibinfo {pages} {8565} (\bibinfo {year} {2003})}\BibitemShut
  {NoStop}%
\bibitem [{\citenamefont {Parker}\ \emph {et~al.}(2013)\citenamefont {Parker},
  \citenamefont {Imberti}, \citenamefont {Callear},\ and\ \citenamefont
  {Albers}}]{Parker1}%
  \BibitemOpen
  \bibfield  {author} {\bibinfo {author} {\bibfnamefont {S.~F.}\ \bibnamefont
  {Parker}}, \bibinfo {author} {\bibfnamefont {S.}~\bibnamefont {Imberti}},
  \bibinfo {author} {\bibfnamefont {S.}~\bibnamefont {Callear}}, \ and\
  \bibinfo {author} {\bibfnamefont {P.}~\bibnamefont {Albers}},\ }\href@noop {}
  {\bibfield  {journal} {\bibinfo  {journal} {Chemical Physics}\ }\textbf
  {\bibinfo {volume} {427}},\ \bibinfo {pages} {44} (\bibinfo {year}
  {2013})}\BibitemShut {NoStop}%
\bibitem [{\citenamefont {Beeman}\ \emph {et~al.}(1984)\citenamefont {Beeman},
  \citenamefont {Silverman}, \citenamefont {Lynds},\ and\ \citenamefont
  {Anderson}}]{Beeman1}%
  \BibitemOpen
  \bibfield  {author} {\bibinfo {author} {\bibfnamefont {D.}~\bibnamefont
  {Beeman}}, \bibinfo {author} {\bibfnamefont {J.}~\bibnamefont {Silverman}},
  \bibinfo {author} {\bibfnamefont {R.}~\bibnamefont {Lynds}}, \ and\ \bibinfo
  {author} {\bibfnamefont {M.~R.}\ \bibnamefont {Anderson}},\ }\href@noop {}
  {\bibfield  {journal} {\bibinfo  {journal} {Phys.Rev.B}\ }\textbf {\bibinfo
  {volume} {30}},\ \bibinfo {pages} {870} (\bibinfo {year} {1984})}\BibitemShut
  {NoStop}%
\bibitem [{\citenamefont {King}(1967)}]{King1}%
  \BibitemOpen
  \bibfield  {author} {\bibinfo {author} {\bibfnamefont {S.}~\bibnamefont
  {King}},\ }\href@noop {} {\bibfield  {journal} {\bibinfo  {journal} {Nature}\
  }\textbf {\bibinfo {volume} {213}},\ \bibinfo {pages} {1112} (\bibinfo {year}
  {1967})}\BibitemShut {NoStop}%
\bibitem [{\citenamefont {Roux}\ and\ \citenamefont {Petkov}(2010)}]{ISSACS1}%
  \BibitemOpen
  \bibfield  {author} {\bibinfo {author} {\bibfnamefont {S.}~\bibnamefont
  {Roux}}\ and\ \bibinfo {author} {\bibfnamefont {V.}~\bibnamefont {Petkov}},\
  }\href@noop {} {\bibfield  {journal} {\bibinfo  {journal} {J. Appl. Cryst.}\
  }\textbf {\bibinfo {volume} {43}},\ \bibinfo {pages} {181} (\bibinfo {year}
  {2010})}\BibitemShut {NoStop}%
\bibitem [{Note4()}]{Note4}%
  \BibitemOpen
  \bibinfo {note} {$\chi ^2$ measures goodness of fit between experimental and
  FEAR model\cite {Anup1,Anup2,Anup3}}\BibitemShut {NoStop}%
\bibitem [{\citenamefont {Drabold}\ \emph {et~al.}(1994)\citenamefont
  {Drabold}, \citenamefont {Fedders},\ and\ \citenamefont {Stumm}}]{DraboldC2}%
  \BibitemOpen
  \bibfield  {author} {\bibinfo {author} {\bibfnamefont {D.~A.}\ \bibnamefont
  {Drabold}}, \bibinfo {author} {\bibfnamefont {P.~A.}\ \bibnamefont
  {Fedders}}, \ and\ \bibinfo {author} {\bibfnamefont {P.}~\bibnamefont
  {Stumm}},\ }\href@noop {} {\bibfield  {journal} {\bibinfo  {journal}
  {Phys.Rev.B}\ }\textbf {\bibinfo {volume} {49}},\ \bibinfo {pages} {16415}
  (\bibinfo {year} {1994})}\BibitemShut {NoStop}%
\bibitem [{\citenamefont {Robertson}\ and\ \citenamefont
  {O'Reilly}(1987)}]{Robertson1}%
  \BibitemOpen
  \bibfield  {author} {\bibinfo {author} {\bibfnamefont {J.}~\bibnamefont
  {Robertson}}\ and\ \bibinfo {author} {\bibfnamefont {E.}~\bibnamefont
  {O'Reilly}},\ }\href@noop {} {\bibfield  {journal} {\bibinfo  {journal}
  {Phys.Rev.B}\ }\textbf {\bibinfo {volume} {35}},\ \bibinfo {pages} {2946}
  (\bibinfo {year} {1987})}\BibitemShut {NoStop}%
\bibitem [{\citenamefont {Chen}\ and\ \citenamefont {Robertson}(1998)}]{Chen1}%
  \BibitemOpen
  \bibfield  {author} {\bibinfo {author} {\bibfnamefont {C.}~\bibnamefont
  {Chen}}\ and\ \bibinfo {author} {\bibfnamefont {J.}~\bibnamefont
  {Robertson}},\ }\href@noop {} {\bibfield  {journal} {\bibinfo  {journal}
  {Journal of Non-Crystalline Solids}\ }\textbf {\bibinfo {volume} {227-230}},\
  \bibinfo {pages} {602} (\bibinfo {year} {1998})}\BibitemShut {NoStop}%
\bibitem [{\citenamefont {Lopinski}\ \emph {et~al.}(1996)\citenamefont
  {Lopinski}, \citenamefont {Merkulov},\ and\ \citenamefont
  {Lanin}}]{Lopinski1}%
  \BibitemOpen
  \bibfield  {author} {\bibinfo {author} {\bibfnamefont {G.~P.}\ \bibnamefont
  {Lopinski}}, \bibinfo {author} {\bibfnamefont {V.~I.}\ \bibnamefont
  {Merkulov}}, \ and\ \bibinfo {author} {\bibfnamefont {J.~S.}\ \bibnamefont
  {Lanin}},\ }\href@noop {} {\bibfield  {journal} {\bibinfo  {journal} {Appl.
  Phys. Lett.}\ }\textbf {\bibinfo {volume} {69}},\ \bibinfo {pages} {3348}
  (\bibinfo {year} {1996})}\BibitemShut {NoStop}%
\bibitem [{\citenamefont {Wang}\ and\ \citenamefont {Ho}(1993)}]{Wang3}%
  \BibitemOpen
  \bibfield  {author} {\bibinfo {author} {\bibfnamefont {C.}~\bibnamefont
  {Wang}}\ and\ \bibinfo {author} {\bibfnamefont {K.}~\bibnamefont {Ho}},\
  }\href@noop {} {\bibfield  {journal} {\bibinfo  {journal} {Phys.Rev.Lett}\
  }\textbf {\bibinfo {volume} {71}},\ \bibinfo {pages} {1184} (\bibinfo {year}
  {1993})}\BibitemShut {NoStop}%
\bibitem [{\citenamefont {Bhattarai}\ and\ \citenamefont
  {Drabold}(2016)}]{Bhattarai1}%
  \BibitemOpen
  \bibfield  {author} {\bibinfo {author} {\bibfnamefont {B.}~\bibnamefont
  {Bhattarai}}\ and\ \bibinfo {author} {\bibfnamefont {D.~A.}\ \bibnamefont
  {Drabold}},\ }\href@noop {} {\bibfield  {journal} {\bibinfo  {journal}
  {Journal of Non-Crystalline Solids}\ }\textbf {\bibinfo {volume} {439}},\
  \bibinfo {pages} {6} (\bibinfo {year} {2016})}\BibitemShut {NoStop}%
\bibitem [{\citenamefont {Li}\ and\ \citenamefont {Drabold}(2013)}]{Yuting1}%
  \BibitemOpen
  \bibfield  {author} {\bibinfo {author} {\bibfnamefont {Y.}~\bibnamefont
  {Li}}\ and\ \bibinfo {author} {\bibfnamefont {D.~A.}\ \bibnamefont
  {Drabold}},\ }\href@noop {} {\bibfield  {journal} {\bibinfo  {journal} {Phys.
  Status Solidi B}\ }\textbf {\bibinfo {volume} {250}},\ \bibinfo {pages}
  {1012} (\bibinfo {year} {2013})}\BibitemShut {NoStop}%
\bibitem [{\citenamefont {Papanek}\ \emph {et~al.}(2001)\citenamefont
  {Papanek}, \citenamefont {Kamitakahara}, \citenamefont {Zhou},\ and\
  \citenamefont {Fischer}}]{Papanek1}%
  \BibitemOpen
  \bibfield  {author} {\bibinfo {author} {\bibfnamefont {P.}~\bibnamefont
  {Papanek}}, \bibinfo {author} {\bibfnamefont {W.~A.}\ \bibnamefont
  {Kamitakahara}}, \bibinfo {author} {\bibfnamefont {P.}~\bibnamefont {Zhou}},
  \ and\ \bibinfo {author} {\bibfnamefont {J.~E.}\ \bibnamefont {Fischer}},\
  }\href@noop {} {\bibfield  {journal} {\bibinfo  {journal} {J. Phys.: Condens.
  Matter}\ }\textbf {\bibinfo {volume} {13}},\ \bibinfo {pages} {8287}
  (\bibinfo {year} {2001})}\BibitemShut {NoStop}%
\bibitem [{\citenamefont {Soler}\ \emph {et~al.}(2002)\citenamefont {Soler},
  \citenamefont {Artacho}, \citenamefont {Gale}, \citenamefont {García},
  \citenamefont {Junquera}, \citenamefont {Ordejon},\ and\ \citenamefont
  {Sanchez-Portal}}]{siesta}%
  \BibitemOpen
  \bibfield  {author} {\bibinfo {author} {\bibfnamefont {J.~M.}\ \bibnamefont
  {Soler}}, \bibinfo {author} {\bibfnamefont {E.}~\bibnamefont {Artacho}},
  \bibinfo {author} {\bibfnamefont {J.~D.}\ \bibnamefont {Gale}}, \bibinfo
  {author} {\bibfnamefont {A.}~\bibnamefont {García}}, \bibinfo {author}
  {\bibfnamefont {J.}~\bibnamefont {Junquera}}, \bibinfo {author}
  {\bibfnamefont {P.}~\bibnamefont {Ordejon}}, \ and\ \bibinfo {author}
  {\bibfnamefont {D.}~\bibnamefont {Sanchez-Portal}},\ }\href@noop {}
  {\bibfield  {journal} {\bibinfo  {journal} {Journal of Physics: Condensed
  Matter}\ }\textbf {\bibinfo {volume} {14}},\ \bibinfo {pages} {2745}
  (\bibinfo {year} {2002})}\BibitemShut {NoStop}%
\bibitem [{\citenamefont {Tucker}\ \emph {et~al.}(2007)\citenamefont {Tucker},
  \citenamefont {Keen}, \citenamefont {Dove}, \citenamefont {Goodwin},\ and\
  \citenamefont {Hui}}]{Tucker2}%
  \BibitemOpen
  \bibfield  {author} {\bibinfo {author} {\bibfnamefont {M.~G.}\ \bibnamefont
  {Tucker}}, \bibinfo {author} {\bibfnamefont {D.~A.}\ \bibnamefont {Keen}},
  \bibinfo {author} {\bibfnamefont {M.~T.}\ \bibnamefont {Dove}}, \bibinfo
  {author} {\bibfnamefont {A.~L.}\ \bibnamefont {Goodwin}}, \ and\ \bibinfo
  {author} {\bibfnamefont {Q.}~\bibnamefont {Hui}},\ }\href@noop {} {\bibfield
  {journal} {\bibinfo  {journal} {J. Phys.: Condens. Matter}\ }\textbf
  {\bibinfo {volume} {19}},\ \bibinfo {pages} {335218} (\bibinfo {year}
  {2007})}\BibitemShut {NoStop}%
\end{thebibliography}%

\end{document}